\documentclass[12pt,preprint]{aastex}

\font\rmsmall=cmr8
\newcommand\litl{\rm\scriptscriptstyle}
\newcommand{\expo}[1]{\ensuremath{10^{#1}}}

\newcommand{\eg}{{\rm e.g.},}
\newcommand{\etal}{{\rm et al.},}

\newcommand{\ie}{{\rm i.e.},}
\newcommand{\kms}{\ensuremath{\,\km\s^{-1}}}
\newcommand{\ciiunit}{\ensuremath{\,\expo{-6}\erg\cmtwo\persec\,{\rm sr}^{-1}}}

\newcommand{\intint}[1]{\ensuremath{\,\times\expo{#1}\,\erg\cmtwo\persec\,{\rm sr}^{-1}}}
\newcommand{\intunit}[1]{\ensuremath{\expo{#1}\,\erg\cmtwo\persec\,{\rm sr}^{-1}}}

\newcommand{\cmtwo}{\ensuremath{\,{\rm cm}^{-2}}}
\newcommand\cmsquared{\ensuremath{\,{\rm cm}^{2}}}
\newcommand\cmthree{\ensuremath{\,{\rm cm}^{-3}}}
\newcommand\K{\ensuremath{\,{\rm K}}}

\newcommand\km{\ensuremath{\,{\rm  km}}}

\newcommand\s{\ensuremath{\,{\rm  s}}}
\newcommand\persec{\ensuremath{\,{\rm  s^{-1}}}}

\newcommand\eV{\ensuremath{\,{\rm  eV}}}
\newcommand\erg{\ensuremath{\,{\rm  erg}}}
\newcommand\mjysr{\ensuremath{\,{\rm MJy}\,{\rm sr}^{-1}}}

\newcommand\hi{H~{\rmsmall I}}
\newcommand\htwo{\ensuremath{{\rm H}_2}}

\newcommand\tex  {\ifmmode{{\rm T_{\rm ex}}}\else{{${\rm T_{\rm ex}}$}}\fi}
\newcommand\tk  {\ifmmode{{\rm T_{\rm k}}}\else{{${\rm T_{\rm k}}$}}\fi}
\newcommand\pp{\ifmmode{{^3{\rm P}_{\litl 1}\rightarrow {^3\rm P}_{\litl 0}}}\else{{$
^3{\rm P}_{\litl 1}\rightarrow {^3\rm P}_{\litl 0}$}}\fi}
\newcommand\ppp{\ifmmode{{^2{\rm P}_{\litl 3/2}\rightarrow {^2\rm P}_{\litl 1/2}}}\else{{$
^2{\rm P}_{\litl 3/2}\rightarrow {^2\rm P}_{\litl 1/2}$}}\fi}
\newcommand\ci{{\rm C}~{\rmsmall I}}
\newcommand\oi{{\rm O}~{\rmsmall I}}
\newcommand\cii{{\rm C}~{\rmsmall II}}

\newcommand\av{\ensuremath{{\rm A}_{\rm v}}}
\newcommand\avc{\ensuremath{{\rm A}_{\rm v,c}}}
\newcommand\nnh{\ifmmode{{{\rm N}_{\litl H}}}\else{{${\rm N}_{\litl H}$}}\fi}
\newcommand\nh{\ifmmode{{{\rm n}_{\litl H}}}\else{{${\rm n}_{\litl H}$}}\fi}
\newcommand\nc{\ifmmode{{{\rm N}_{\litl C}}}\else{{${\rm N}_{\litl C}$}}\fi}
\newcommand\ncii{\ifmmode{{{\rm N}_{\litl CII}}}\else{{${\rm N}_{\litl CII}$}}\fi}
\newcommand\nco{\ifmmode{{{\rm N}_{\litl CO}}}\else{{${\rm N}_{\litl CO}$}}\fi}
\newcommand\ncodv{\ifmmode{{{\rm N}_{\litl CO}/{\Delta{\rm V}}}}\else{{${\rm N}_{\litl CO}/{\Delta{\rm V}}$}}\fi}
\newcommand\xco{\ifmmode{{{\rm X}_{\litl CO}}}\else{{${\rm X}_{\litl CO}$}}\fi}
\newcommand\ncrit{\ensuremath{{\rm n}_{\litl crit}}}

\newcommand\irasd{\ensuremath{{\rm I}_{\litl 100}}}
\newcommand\irasc{\ensuremath{{\rm I}_{\litl 60}}}
\newcommand\icii{\ensuremath{{\rm I}_{\litl CII}}}

\slugcomment{To appear in the Astrophysical Journal, November 2002}

\shorttitle{Heating and Cooling in High Latitude Clouds}
\shortauthors{Ingalls, Reach, and Bania}

\begin{document}


\title{Photoelectric Heating and [\cii ] Cooling of High 
Galactic Latitude Translucent Clouds\footnote{Based on observations with ISO, 
an ESA project with instruments funded by ESA Member
States (especially the PI countries: France, Germany, the Netherlands 
and the United Kingdom) with the participation of ISAS and NASA.}}


\author{James G. Ingalls and William T. Reach}
\affil{SIRTF Science Center; California Institute of Technology, 1200 E California Blvd, Mail Stop 220-6, Pasadena CA 91125;\\
ingalls@ipac.caltech.edu,reach@ipac.caltech.edu }
\and
\author{T. M. Bania}
\affil{Institute for Astrophysical Research; Boston University; 725 Commonwealth Avenue,
Boston, MA 02215;\\ bania@bu.edu}


\begin{abstract}
The (\ppp ) transition of singly--ionized carbon, [\cii ], is 
the primary coolant of diffuse interstellar gas.  We describe
observations of [\cii ] emission towards nine high Galactic latitude
translucent molecular clouds, made with the long wavelength spectrometer
on board the Infrared Space Observatory.  To understand the role of 
dust grains in processing the interstellar radiation 
field (ISRF) and heating the gas, we compare the [\cii ] integrated 
intensity with the 
far-infrared (FIR) integrated surface brightness for the 101 sampled lines 
of sight.  We find that [\cii] is linearly correlated with FIR, and 
the average ratio is equal to that measured with the {\sl COBE} satellite
for all high-latitude Milky Way gas.  There
is a significant decrease that was not detected with {\sl COBE}
in [\cii ] emissivity at high values of FIR.
Our sample splits naturally into two populations depending on the 
60\micron/100\micron\ surface brightness ratio, or color:  ``warm'' positions 
with $60/100 > 0.16$, and ``cold'' positions with $60/100 < 0.16$.  
A transition from sources with warm to those with cold 60/100
colors coincides approximately with the transition from constant to
decreasing [\cii ] emissivity.  We model the [\cii ] and 
far--infrared emission under conditions of thermal equilibrium, using
the simplifying assumptions that, in all regions heated by the ISRF, the 
most important source of gas heating is the photoelectric 
effect on grains and the most 
important source of gas cooling is [\cii ] emission.  The model matches
the data well, provided the ISRF incident flux is $\chi_0 \approx 1.6$ 
(in units of the nominal value near the Sun), and the photoelectric heating 
efficiency is $\epsilon \approx 4.3\%$.  There are no statistically
significant differences in the derived values of $\chi_0$ and 
$\epsilon$ for warm and cold
sources.  The observed variations in the [\cii ] emissivity and
the 60/100 colors can be understood entirely in terms of
the attenuation and softening of the ISRF by translucent clouds, not
changes in dust properties.
\end{abstract}




\section{Introduction}
Diffuse interstellar gas clouds are heated by the absorption of starlight 
and cooled by the emission of spectral line radiation at far--infrared
(FIR) and submillimeter wavelengths.  Energy can be 
transferred from the interstellar radiation field (ISRF) to the gas by the
 photoelectric 
effect on dust grains \citep{wat72,dej77,dra78,bak94},
whereby far ultraviolet (FUV) (6\,eV $< h\nu <$ 13.6\,eV) photons ionize 
grains and the ejected electrons heat 
the gas through collisions.  In diffuse \hi\ regions of low kinetic 
temperature ($\tk \sim 50-200\K$), the ``cold neutral medium'' (CNM),
this process is thought to be the dominant gas heating mechanism
\citep[see][]{wol95}.  

The most abundant form of carbon gas in the CNM is the singly-ionized carbon
atom, \cii, because the first ionization potential of carbon is 11.3\,eV and
the CNM is permeated by FUV photons.  The 158\micron\ (\ppp ) emission line 
transition of [\cii ] should be the primary coolant of the CNM 
\citep{wol95}.  There are two main reasons for this.  First, 
the \cii\ ion is relatively easy to excite by 
collisions given average CNM temperatures ($T\sim 80\K$;\citealt{kul87}).  
After the hyperfine 
states of hydrogen, the $^2P_{3/2}$ fine structure 
state of \cii\ is the first excited state ($h\nu/k = 91\K$) of all gas-phase 
CNM constituents (the next state is the $^3P_1$ level of the \oi\ atom, which
is 228\K\ above ground).  Second, once excited, the energy loss
rate, given by the product $h\nu\,A_{ul}\times \ncii[^2P_{3/2}]$, is 
much higher for \cii\ than for other CNM species.  Here $A_{ul}$ is 
the Einstein spontaneous
emission coefficient for the \ppp\ transition and $\ncii[^2P_{3/2}]$ 
is the column density of \cii\ in the excited state .

Since the FUV radiation field that ionizes dust grains (and thus heats the 
gas) can also heat the grains, 
one expects a connection between the [\cii ] 158\micron\ line emission from 
photoelectron-heated gas and the far-infrared (FIR) emission 
from FUV-heated dust grains.  Large-scale surveys of the Galaxy indeed 
reveal a link between the \cii\ and FIR emission properties of the CNM.  
An unbiased survey of spectral line and continuum 
emission in the Milky Way was conducted with two instruments on board the 
Cosmic Background Explorer ({\sl COBE}) satellite:  the Far Infrared 
Absolute Spectrometer (FIRAS) and the Diffuse Infrared Background Experiment 
(DIRBE).  The [\cii ] emission was observed with FIRAS to
be closely correlated with \hi\ emission at high Galactic latitudes, with
an average \cii\ cooling rate of $(2.65\pm 0.15)\times 
10^{-26}\,$erg\,s$^{-1}\,$(H~atom)$^{-1}$ \citep{ben94}.  Both 
DIRBE and FIRAS maps of the dust continuum show that the dust 
opacity and FIR surface brightness are also correlated with \hi\ column 
density at high latitude \citep{bou96}. 

So on the large scale (FIRAS has an angular resolution of 7\arcdeg, DIRBE of 
40\arcmin), CNM \hi\ gas has constant FIR {\it and} [\cii ]
emissivity.  It is therefore likely that the processing of FUV radiation 
by dust grains (1) to produce FIR dust emission and (2) to heat the 
gas---resulting in [\cii ] emission---occurs in similar ways throughout 
the cold interstellar 
medium.  It is difficult to test this idea directly using the {\sl COBE} 
data, mainly because of inadequate resolution.  The emission from 
gas with 
vastly different physical conditions, for example molecular and atomic 
gas, might be combined together in a single
FIRAS or DIRBE measurement, making it impossible to understand the physical
mechanisms responsible for the observed emission.  

High resolution observations of the [\cii ] and FIR emission from translucent 
molecular clouds can be used to test the limits of models for gas heating by 
the ISRF.  Molecular gas can only exist in shielded locales where the FUV 
portion of the ISRF has been attenuated, and thus where the photoelectric 
heating is much weaker than in atomic regions.  In these regions, most 
of the carbon has combined into CO, and most of the [\cii ] emission has 
``turned off.''  Although FUV radiation is weak in molecular zones,  
optical photons are still present to heat dust grains and produce FIR 
emission.  Indeed, many {\sl IRAS} cirrus 
clouds \citep{low84} first detected in FIR emission at 60 and
100\micron\ are associated with intermediate-extinction molecular clouds, or 
translucent clouds \citep{van88}.   Translucent 
molecular clouds are most easily observed when nearby
and at high Galactic latitudes ($b\gtrsim 15\arcdeg$) \citep{mag85}.  Currently over 100 
nearby ($\langle d\rangle \approx 105\,$pc) translucent high--latitude clouds 
(HLCs), have been cataloged using their CO emission \citep*[see][]{mag96}.  In a study of 75 HLCs, it was shown that all
of the clouds are associated with \hi, and that the HLCs probably condensed 
out of diffuse CNM gas \citep*{gir94}.

We examine here the [\cii ] and FIR emission 
towards a sample of translucent HLCs using the {\sl ISO} satellite.  Since 
the mean angular size of HLCs in CO emission maps is $\sim 1\arcdeg$ 
\citep{mag96}, the clouds are easily resolved by the
$\sim 71\arcsec$ beam of {\sl ISO}.  This allows us to choose
between lines of sight towards individual clouds with varying amounts of 
atomic and molecular gas, 
which was not possible with {\sl COBE}.  As described above, the lines of 
sight dominated by
molecular gas include regions from which FUV photons are excluded.    

In this paper we summarize the {\sl ISO} observations of
[\cii ] emission towards HLCs, all of which are available from the {\sl ISO} 
archive (\S2).  We discover that HLCs fall into two categories based on their
60\micron/100\micron\ color, ``cold'' sources and ``warm'' sources.  
We find that the line-of-sight [\cii ] emissivity is consistent
with the average high--latitude {\sl COBE} cloud emission for both
cold and warm positions.  We surmise that the chief distinction
between the two 60/100 color regimes is the column density (\S3).  Assuming 
that the local heating and cooling depend only on the fixed
properties of dust grains and the attenuated ISRF intensity, we develop a
model for the [\cii ] and FIR emission of translucent clouds.  The 
model enables us to estimate the average line of sight
photoelectric heating efficiency, $\epsilon$, as well as the intensity of the 
interstellar radiation field incident on HLCs, $\chi_0$ (\S4).  We discuss the
significance of our results, and point out some issues unresolved by this 
study (\S5).  Finally, we summarize our conclusions (\S6).

\section{Observations}

\subsection{ISO Observations of Translucent High--Latitude Clouds}
The Infrared Space Observatory ({\sl ISO}) \citep{kes96}, a satellite 
observatory built by the European Space Agency
(ESA), was launched in 1995 November and operated until its cryogenic
reserves were depleted in 1998 April.  A total of seven peer-reviewed 
observing programs dedicated to studying 
[\cii ] emission from nine translucent high--latitude molecular clouds were 
carried out with the long wavelength spectrometer (LWS) \citep{cle96} 
on board {\sl ISO}.  Some of these [\cii ] measurements 
were the results of proposals by us (observer:proposal = {\tt TBANIA:TMB\_1},
{\tt TBANIA:TMB\_2}, {\tt TBANIA:SH\_HLC}, and {\tt WREACH:PHASE}), and the 
rest were 
extracted from the {\sl ISO} public data archive.\footnote{See the web site 
{\it http://www.iso.vilspa.esa.es/} for details.}  The {\tt TBANIA} data
were originally published in \citet{ing99}.  The data for twenty of the
positions observed towards cloud MBM-12 (L1457) were reported previously
by \citet{tim98}.

Table 1 lists the nine HLCs observed in [\cii ] with {\sl ISO}, together 
with their positions in Galactic coordinates ($\ell, b$) (Columns 2 and 3).  
Each entry in the Table represents a separate source
``observation,'' \ie\ a self-contained set of measurements of the target.  A 
given observation may consist of a single point, or it may consist of multiple 
pointings of the telescope as part of a raster map.  For clouds with
$|b| > 25\arcdeg$ (all clouds in our sample but G300.1--16.6) we use
the cloud names given in the \citet{mag96} catalog, and list commonly used 
alternate names in parenthesis (column 1
of Table 1).  The field of view (FOV) of each observation (column 4) is 
defined by the full width at half maximum (FWHM) beam size of the {\sl ISO}
LWS instrument as well as the total extent of the raster grid employed.  
We indicate the position angle (PA) of raster grids, measured counterclockwise
from celestial North, in column 5.

Emission spectra of [\cii ] were measured with the LWS spectrometer in 
either LWS01 or LWS02 scanned grating modes.  Rapid scanning was used,
and each spectral resolution element 
(of width $\sim 0.6\micron$) was sampled many times.  This
ensured that cosmic ray hits on the detector were efficiently diagnosed
and removed.  We used detector LW4, centered on 161$\,\mu$m, to measure the 
157.7\micron\ (\ppp ) transition of [\cii ].  This detector is
estimated to have a nearly circular beam, with FWHM size $\sim 71\arcsec$ 
\citep{swi98}, 20\% smaller than predicted by early optical 
models.  The
truncation of the LWS beam is attributed to multiple clipping effects 
and stray light propagation within the optical system 
\citep{cal98}.  Stray light is also apparently the cause of 
interference fringes, which manifest as a sinusoidal modulation with 
period $\sim 10\micron$ at $\lambda = 158\micron$ in the LWS spectra of 
extended sources \citep*{tra96}.  We
attempted to minimize the effect of the fringes on our [\cii ] integrated
intensities by subtracting either
a linear or quadratic baseline from the spectral data measured with the LW4
detector.

We derived \cii\ integrated intensities using
the Automatic Analysis (AA) data products from the {\sl ISO}
archive.  All calibration steps were performed at ESA in batch mode, 
\ie\ without human input.  In order to check the quality of the AA procedure,
we calibrated a subset of our spectra manually using the LWS Interactive 
Analysis software and 
compared the results with the the corresponding AA results.  The 
integrated [\cii ] 
intensities produced using the two methods were indistinguishable to within 
$\pm 10\%$.  As the absolute flux calibration of the LWS has been estimated
to be no better than $\pm 20\%$ \citep{bur98}, we judged that
calibration beyond the AA stage was not necessary.

We reduced the calibrated scan data using the ISO Spectral Analysis Package
(ISAP).\footnote{The ISO Spectral Analysis Package (ISAP) is a joint 
development by the LWS and SWS Instrument Teams and Data Centers.
Contributing institutes are CESR, IAS, IPAC, MPE, RAL and SRON.}  First, we 
averaged all scans and scan directions for detector LW4 in bins of width 
0.2\micron\ using a 3$\sigma$ median clipping algorithm.  Then we subtracted 
linear or quadratic baselines from the continuum, and made Gaussian fits 
to each spectrum.  Since the Doppler equivalent velocity 
resolution of the spectra
is $\sim 1000\kms$, which is much larger than the typical \hi\ linewidth of 
HLCs (6-26\kms), we assumed that all measurements were unresolved 
spectrally.  This
allowed us to fix the FWHM linewidth in the Gaussian fits to the known LWS
resolution of $0.6\micron$.  We used the Gaussian fits to compute the [\cii ] 
integrated flux, which we divided by
the LWS beam area, $\Omega = 9.3\times 10^{-8}\,$sr, to obtain the [\cii ]
integrated intensity, \icii, in units of erg$\cmtwo\persec\,{\rm sr}^{-1}$.  
We derived error bars in \icii\ by adding a 20\% absolute calibration 
uncertainty \citep{bur98} in quadrature with the formal Gaussian 
fit errors in the [\cii ] line integral, obtained using the ISAP software.

\subsection{IRAS 60 and 100\micron\ Measurements}

We obtained {\sl IRAS} Sky Survey Atlas (ISSA) images of each cloud and 
extracted the 60\micron\ and 100\micron\ surface brightness, \irasc\ and 
\irasd\ respectively, for the positions observed in [\cii ].  The 
angular resolution of the ISSA is only 4\arcmin, whereas the {\sl ISO} 
LWS has a FWHM beamsize of 71\arcsec\ at 157.7\micron, so
we interpolated the {\sl ISSA} maps 
onto a 71\arcsec\ grid before making \irasc\ and \irasd\ measurements.  

We have tested the validity of interpolating the 4\arcmin\ {\sl IRAS} images
to compare with {\sl ISO} data.  We obtained
HIRES-processed {\sl IRAS} images of the clouds in our 
sample \citep{lev93}.  The HIRES maps have approximately 60\arcsec\ and 
100\arcsec\ resolution at 60 and 100\micron, 
respectively, which is comparable to the {\sl ISO} LWS beamsize.  A
linear fit to the HIRES data as a function of interpolated ISSA 
surface brightness
shows a strong correlation with slope=1 for pixels
with ISSA values brighter than about half of the maximum value.  Below 
the half-maximum threshold, HIRES
images have an average surface brightness of zero, probably the result of
background subtraction in the HIRES process.  We conclude that on 
average the interpolated ISSA images give a good approximation to the 
actual surface brightness at
higher resolution, with the advantage of maintaining the low
surface brightness emission that does not appear in HIRES images.

The ISSA plates are optimized for relative, not 
absolute, photometry \citep{whe94}.  To put images of different
clouds on the same relative scale, we devised a zeropoint 
adjustment method that took advantage of the constant
60\micron/100\micron\
scaling properties of high--latitude clouds.  Pixel-by-pixel
comparisons of the 60\micron\ and 100\micron\ images show a strong 
correlation (the correlation coefficient is typically greater than 0.9).  
Linear fits to the \irasd\
versus \irasc\ data for each of our nine clouds yield the average 
relationship \irasd = $(5.74\pm 0.15)$\irasc
-- $(1.14\pm 0.13)$.  The relative variation in the derived slopes is
0.15/5.74$\sim3\%$, showing that the 60/100 {\it ratio} is nearly
constant for all cloud images.  On the other hand, the relative variation 
in the \irasd -intercepts exceeds $11\%$, indicating significant 
fluctuations in the 60\micron\ and 100\micron\ surface brightness zeropoint
from image to image.   We adjusted the zeropoint of each image so that
 \irasd =0 when \irasc =0 by subtracting
the derived \irasd -intercept values from each 100\micron\ cloud image.

We converted the
{\sl IRAS} relative scale to the {\sl COBE} DIRBE absolute surface 
brightness scale by multiplying the 60\micron\ data by 0.87 and the 
100\micron\ data by 0.72 \citep{dir98}.  We derived error 
bars in \irasd\ by combining in quadrature the 10\%
accuracy in point source flux recovery \citep{ira88},
with the 7\% uncertainty in the DIRBE to {\sl IRAS} 
conversion at 100\micron.  We derived error bars for \irasc\ measurements 
by combining in quadrature
the 10\% accuracy in point source flux recovery with the 2.5\% 
uncertainty in the DIRBE to {\sl IRAS} conversion at 60\micron.

\section{Observational Results}

A total of 109 positions in HLCs were observed using the {\sl ISO} LWS, and 
[\cii ] emission was detected towards 101 of them.  We 
list the [\cii ] detection rate for each observation in column 6 of 
Table 1.  

\subsection{The HLCs as Representative [\cii ]  Clouds}
\citet{ing00} showed that HLCs are the average Galactic high-latitude 
sources of CO and far-infrared emission.  We demonstrate here 
that they are also representative sources of [\cii ] radiation.  We 
mentioned in \S1 that CNM \hi\ emission is correlated with both 
far-infrared and [\cii ] emission.  Now we compare directly the $FIR$ 
and [\cii ] properties of high-latitude material using the
all sky {\sl COBE} surveys.  The FIRAS Line Emission Maps 
give [\cii ] 157.7\micron\ integrated intensity at 7\arcdeg\ spatial 
resolution.  The DIRBE Zodi-Subtracted
Mission Average (ZSMA) maps give dust continuum surface brightness in
far-infrared bands at 40\arcmin\ resolution \citep{dir98}.  
We use a function introduced by \citet{hel85} \citep[see also][]{hel88} to 
define the far-infrared 
surface brightness in terms of the 60 and 100\micron\ surface brightness:
\begin{equation}
\frac{FIR}{\intunit{-6}} = 12.6 \left( \frac{\irasd + 2.58\,\irasc}{\mjysr} \right)
.\label{fir_def}
\end{equation}
This representation of far-infrared emission is estimated to be accurate 
to within 1\% for modified blackbody spectra from dust clouds with 
temperatures between 20 and 80\,K and emissivity that varies as
$\nu^{0-2}$.  

\begin{figure}
\plotone{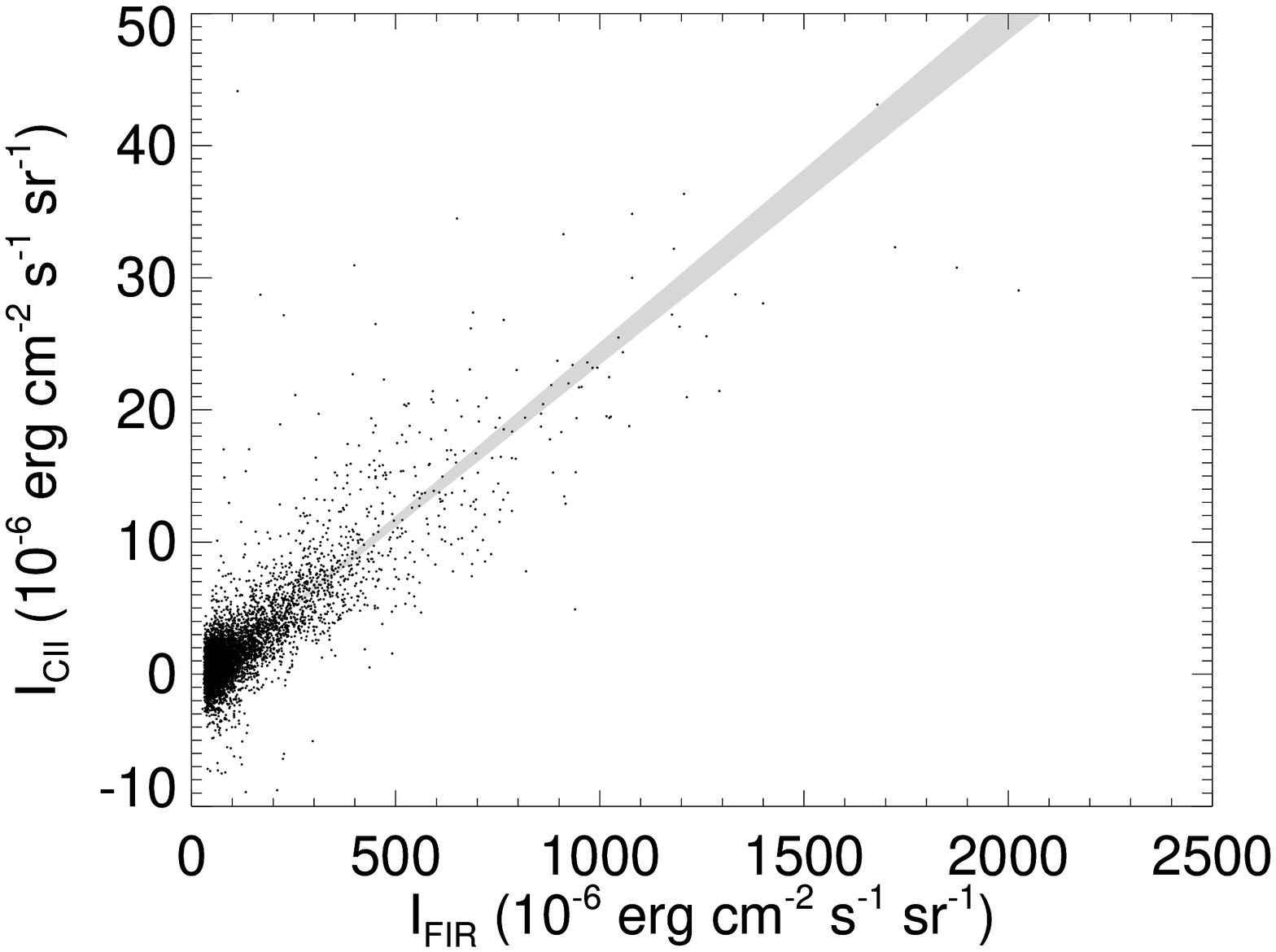}
\caption{High Galactic latitude
[\cii ] emission is correlated linearly with $FIR$ emission on the
large scale.  We plot here the integrated intensity of [\cii ] (\ppp ) 
emission from 
{\sl COBE} FIRAS Line Emission Maps, \icii, as a function of 
the integrated $FIR$ surface brightness, derived from {\sl COBE} DIRBE ZSMA 
60 and 100\micron\ maps resampled to the $7\arcdeg$ FIRAS grid.  The
5523 data points with $|b|>5\arcdeg$ can be fit by a straight line 
with slope $(2.54\pm 0.03)\times\expo{-2}$.  We superimpose a gray 
wedge on the data that represents the $\pm 3\sigma$
range of expected ($FIR$, \icii ) pairs. \label{cii_fir_cobe}}
\end{figure}

We created DIRBE maps of $FIR$ using the 60 and 100\micron\ 
ZSMA data, resampled to the 7\arcdeg\ FIRAS grid, and compared with
the [\cii ] map (see Figure \ref{cii_fir_cobe}).  We obtain a highly 
significant correlation (there is less 
than 0.05\% chance that the data are uncorrelated).  A linear fit weighted 
by the error bars for all data with $|b| > 5\arcdeg$ gives 
\begin{equation}
\icii = (2.54\pm 0.03)\times\expo{-2} FIR -(1.1375\pm 0.0004)\times\ciiunit,
\label{cp_fir_eq}
\end{equation}
where $\icii$ and $FIR$ are in the same units.  A grey wedge on Figure
\ref{cii_fir_cobe} shows the $3\sigma$ range of expected $FIR$ and 
\icii values.  Since the fit was weighted by the {\sl COBE} error bars, 
this wedge is much narrower than the actual scatter in data points (the rms 
scatter in \icii\ about the fit is $5.29\times\ciiunit$).  The fit gave an 
$\icii$-intercept of $-1.1375\times\ciiunit$, because 
many of the FIRAS [\cii ] values for 
low $FIR$ are negative.  For what follows, we assume that in the 
absence of $FIR$ emission there is no [\cii ] emission, and ignore the
intercept.  

\begin{figure}
\plotone{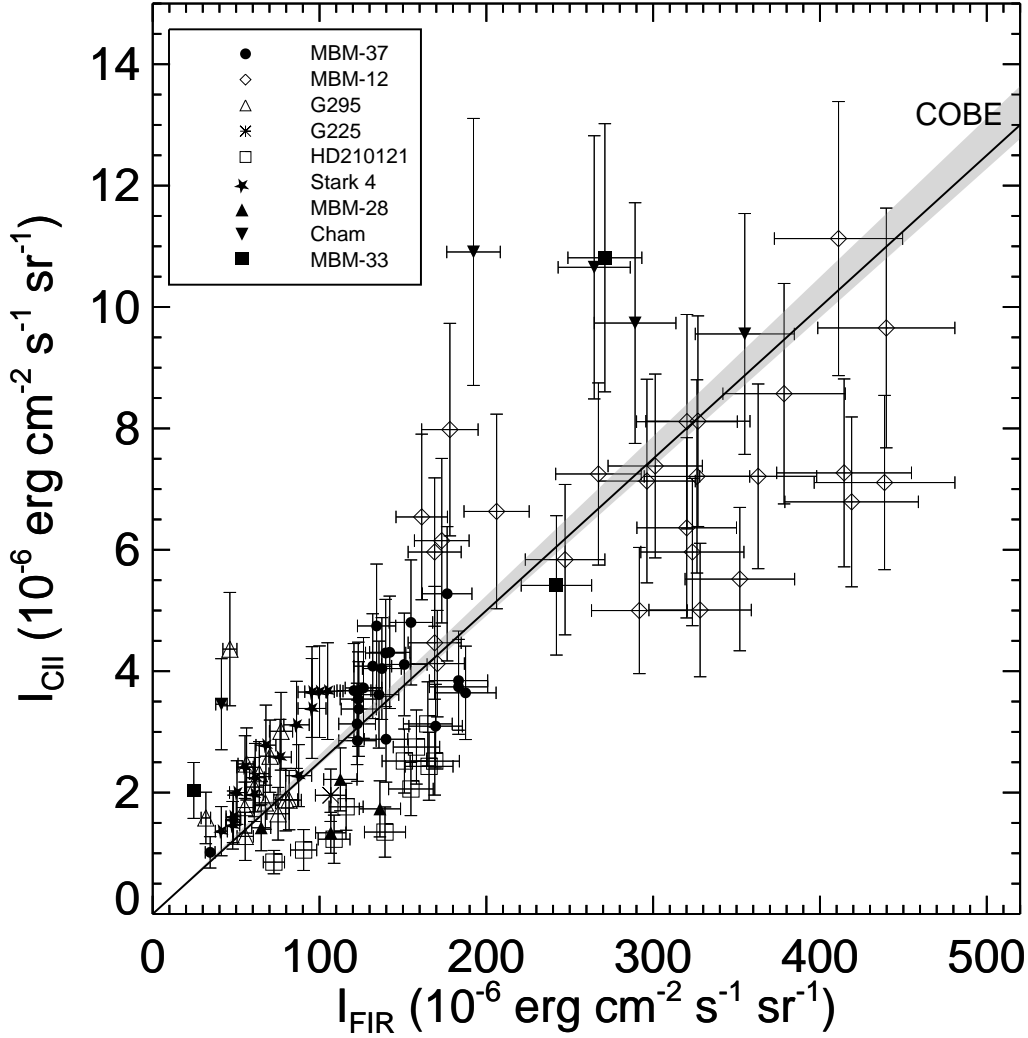}
\caption{The HLCs are average sources of [\cii ] and FIR emission.  
Data points show the {\sl ISO} integrated intensity of [\cii ] 
versus $FIR$ from {\sl IRAS} observations for 101 positions in 9 
high--latitude clouds.  The gray wedge labeled ``COBE'' represents the 
$\pm 3\sigma$ range of expected ($FIR$, \icii ) pairs based on fits to the 
all sky {\sl COBE} data for $|b|>5\arcdeg$ (see Figure \ref{cii_fir_cobe}). 
The solid line is a weighted least squares fit to our HLC data, 
$\icii = 2.5 \times\expo{-2} FIR$.  \label{cp_fir}}  
\end{figure}

We compare the [\cii]-$FIR$ relationship for ``COBE clouds'' with
our HLC data.  The \cii\ integrated intensities of the 101 HLC positions 
detected with 
{\sl ISO} are plotted as a function of the $FIR$ surface brightness 
in Figure \ref{cp_fir}.  Linear regression to the data, weighted by the 
errors, gives 
\begin{equation}
\icii = (2.5 \pm 0.9)\times\expo{-2} FIR.  
\label{cp_fir_fit}
\end{equation}
In the fit we have fixed the \icii -intercept at zero to reflect our 
assumption that $\icii=0$ when $FIR=0$.  
We plot the fit as a solid line in Figure \ref{cp_fir}.  
A gray wedge labeled ``COBE'' represents
the $\pm 3\sigma$ range of expected ($FIR$, \icii) values based on
the {\sl COBE} data (Equation \ref{cp_fir_eq}).  The HLC and 
{\sl COBE} slopes are identical.  {\it Our analysis 
is consistent with the HLCs being part of the same population 
of sources as the average CNM source observed with the low-resolution 
{\sl COBE} instruments}.  

There is evidence that at the {\sl ISO} resolution a straight line is 
not the best representation of the $\icii$-$FIR$ data.  Breaking 
the data into $FIR$ subregions and performing separate linear 
fits yields a systematic decrease in slope as $FIR$ increases.  Figure 
\ref{cp_fir_bins} shows an example of this for bins of width 
$\expo{-4}\erg\cmtwo\persec\,{\rm sr}^{-1}$ along the $FIR$-axis.  
  The local slope is close to the overall slope for 
$FIR\lesssim 2\intint{-4}$, but is virtually zero for 
$FIR\gtrsim 2\intint{-4}$.  We also plot in 
Figure \ref{cp_fir_bins} the weighted mean values of $\icii$ and $FIR$
in the bins.  A significant departure from the 
{\sl COBE} and {\sl ISO} predictions is seen in the bin with 
the largest $FIR$ integrated surface brightness, where 
\icii\ is 25\% lower than expected.  Thus, {\it the [\cii ] emissivity
of HLC gas decreases as $FIR$ increases}.  In \S4 we explain this 
using a model of the radiative heating and cooling of translucent clouds.

\subsection{The 60/100 color}
\citet{ing00} showed that the eight HLCs they studied have typical 60/100 
colors (\irasc/\irasd) when compared to {\sl COBE} clouds.  The
60/100 colors for our nine clouds are likewise typical.  We plot \irasd\ 
as a function of \irasc\ for our sample 
in Figure \ref{iras_100_60}.  Superimposed on the plot is a dashed line 
representing the average behavior for Milky Way gas correlated with \hi\
emission, $\irasc = 0.16\,\irasd$ \citep{dwe97}.  Most positions 
fall slightly below this line.  These sources
probably contain mostly atomic gas, since their $FIR$ surface 
brightnesses are low and their 
\irasc/\irasd\ ratios are slightly ``warmer'' than that
of average \hi.  For $\irasd > 4\mjysr$ there is a split into two 
populations:  the ``cold'' sources with 
$\irasc < 0.16\,\irasd$; and the ``warm'' sources with 
$\irasc \gtrsim 0.16\,\irasd$.  

Warm positions generally produce
less emission than cold positions---over 90\% of the warm positions, but
less than 30\% of the cold positions, have 
$FIR<2\times\expo{-4}\erg\cmtwo\persec\,{\rm sr}^{-1}$.  
Their [\cii] and $FIR$ properties are not easily distinguishable otherwise.
Weighted linear fits to the [\cii] vs. $FIR$ data, holding the intercept 
fixed at zero, give slopes of $(2.6\pm 1.1)\times\expo{-2}$ for warm
sources and $(2.2\pm 1.7)\times\expo{-2}$ for cold sources.  These
slopes are equivalent within the errors to each other, and to the 
slope determined for the sample as a whole.  

The most striking difference between cold and warm positions seems 
to be their {\it column density}.  Figure 
\ref{mbm12_av_color} shows this for
high--latitude cloud MBM--12 (L1457).  Here we compare the 60 and 100\micron\ 
images to a visual extinction map of the cloud.  The extinction, \av, was
derived using an adaptive grid star count algorithm 
\citep{cam97} from H and K images in the 
2MASS second incremental data release.\footnote{This publication makes use of data products from the Two 
Micron All Sky Survey, which is a joint project of the University of 
Massachusetts and the Infrared Processing and Analysis Center/California 
Institute of Technology, funded by the National Aeronautics and Space
Administration and the National Science Foundation.} 
  After resampling the 60\micron, 100\micron, and
\av\ images onto the same grid, we compared the 60/100 color with \av\ 
for each pixel.  Figure \ref{mbm12_av_color} shows 
\irasc/\irasd\ averaged in
bins of width $\Delta\av= 0.71\,$mag.  The 60/100 color decreases 
monotonically with increasing extinction, implying that warm dust 
emission is a characteristic of low extinction positions, while cold 
dust emission is a characteristic of high extinction positions.  
The effect is not slight.  A small adjustment in 60/100, \eg\ from 
0.184 to 0.160, represents a large change in \av, from approximately 1 to 3.  

In what follows, we develop
a model for the radiative transfer of the ISRF in translucent clouds.  This
allows us to predict the emission from dust grains, as well as the 
photoelectric heating 
and [\cii ] cooling of the gas.  We show that the cold/warm
dichotomy in translucent clouds is probably caused by the attenuation of the 
ISRF, not changes in dust properties.

\section{Modeling the \cii\ and FIR emission of HLCs}
The [\cii ] emissivity towards translucent HLCs
decreases as the $FIR$ surface brightness increases.  The 
cold and warm populations of HLC positions, identified 
via their 60/100 colors, have different column 
densities, and hence different amounts of atomic versus molecular gas,
but similar [\cii ]-$FIR$ emission properties.
To explain these empirical facts we offer a model of the heating 
and cooling of gas and dust in the interstellar medium.  The model 
describes regions heated primarily by the interstellar 
radiation field, \ie\ translucent regions.  We simplify the calculations
by treating only the single most important sources of heating and cooling of 
translucent gas:  the photoelectric 
effect for gas heating, and [\cii ] emission for gas 
cooling \citep{wol95}.  We predict the $FIR$ and [\cii ] 
emission under conditions of thermal equilibrium.   For the dust grains, this 
requires that the power emitted by each grain equals the power 
absorbed from the attenuated interstellar radiation field.  For the gas 
this requires that the local [\cii ] cooling rate equals the photoelectric 
heating rate.  

By assuming that [\cii ] emission is tied to photoelectric heating, we
bypass chemistry.  We do not attempt to determine 
the {\it abundance} of
\cii\ gas in the [\cii ]-emitting regions, \eg\ by balancing the production
and destruction of \cii\ using a chemical model.  Instead we determine the
photoelectric heating, which depends on the dust properties, and equate it
with the [\cii ] cooling modified by an efficiency parameter
\citep[e.g., see][]{bak94}.  This 
approximation assumes that (1) most [\cii ] cooling takes place
in FUV-heated areas and (2) most FUV heating takes place in areas 
cooled by [\cii ] emission.  

If \cii\ were abundant in regions not heated by the 
photoelectric effect, then much of the [\cii ] emission might be the
result of another heating mechanism, such as cosmic rays, and we would
underestimate \icii.  This is unlikely.  If 
the radiation field is too weak to ionize dust grains 
(average work 
function:  5.5\eV), it is also too weak to ionize \ci\ (ionization 
potential:  11.3\eV) and form \cii.  In other words, {\it where there 
is [\cii ] cooling, it is the result of FUV heating}.  

The second tier of our approximation maintains that most
photoelectric heating occurs in regions cooled by [\cii ] emission.  
If this did not hold, then the emission
from another species like \ci\ or CO might provide most of the cooling
in some FUV-heated regions and our \icii\ predictions would be 
overestimates.  It is certainly true that \ci\ and CO are important 
coolants of translucent clouds, but in terms of total flux the [\ci ] and
CO lines are much weaker than the [\cii ] line.  The 
brightest known HLC source of [\ci ],
MBM--12, has a maximum observed [\ci ] (\pp ) integrated intensity of about
$3.6\times\expo{-7}\,{\rm erg} \cmtwo\persec\,{\rm sr}^{-1}$ \citep*{ing94}, 
a factor of 10 weaker than the [\cii ] emission towards
most of our HLCs.  According to the photodissociation region models of
\citet{kau99}, the other [\ci ] fine structure transition, 
($^3P_2\rightarrow ^3P_1$), should have at most twice the power
of the (\pp ) transition under translucent conditions (FUV flux equal 
to the local interstellar value and gas volume density 
$\sim \expo{2-5}\cmthree$).  Even if 
all of the \ci\ atoms that emit in these transitions 
were excited by FUV-ejected electrons, the photoelectric heating in the
[\ci ]-emitting regions of HLCs is still at most only 30\% of that in the
[\cii ]-emitting areas.  Furthermore, CO cooling is almost
an order of magnitude weaker than [\ci ] cooling for translucent clouds
\citep[eg., see][]{ing00}.  Thus, {\it where there is significant FUV heating,
most of the cooling is by [\cii ] radiation}.

\subsection{The penetration of the interstellar radiation field into 
translucent clouds}
Here we calculate the intensity 
spectrum of the interstellar radiation field inside a model cloud, as it 
is attenuated and processed by dust grains.  We adopt a plane
parallel geometry, where changes in the mean radiative properties of the 
gas and dust are dependent only on the optical depth, $\tau_{\lambda}$,
measured from the surface of the cloud to the center.  The cloud size 
is defined by the central optical depth, $\tau_{c,\lambda}$.

The interstellar radiation field, $J_{{\litl ISRF},\lambda}$, is isotropic and 
incident on both sides of our plane parallel clouds.  We consider only the 
stellar component of the ISRF longward of 0.0912\micron\ (the Lyman 
continuum), and the cosmic background radiation field.  We use the 
ISRF approximation of \citet*{mez82} and \citet*{mat83}, which is the sum of a 
stellar UV component and three diluted blackbodies.  We add to this the cosmic 
background radiation field, $B_{\lambda}(2.7\K)$, where $B_{\lambda}(T)$
is the Planck function.  We 
calculate the integrated intensity of the interstellar radiation field to be
\begin{equation}
\int_{0.0912\micron}^{1000\micron} 4\pi J_{{\litl ISRF},\lambda} d\lambda = 
2.76\times\expo{-2}\,{\rm erg}\cmtwo\persec.
\label{isrf_integrated}
\end{equation}
This is 27\% higher than the value quoted by \citet{mat83}, partly
because we consider a larger range of wavelength than they did.  In
constructing an ISRF, we do not include clouds themselves as infrared 
sources.  Infrared photons have
little affect on the energetics of translucent clouds because the clouds
are transparent to them.  Furthermore, the interstellar infrared spectrum
results from the processing of stellar radiation by clouds,
and does not represent a new source of energy.  A goal of this paper is to 
calculate the infrared spectrum of translucent clouds, to compare with 
observations.

We use the technique of \citet*{fla80} to model
the penetration of the ISRF into interstellar clouds.  They expressed 
analytically the solution of the radiative transfer equation in the presence 
of scattering using a spherical harmonics method.  
The mean intensity of the radiation field at depth $\tau_{\lambda}$,
for a cloud of size $\tau_{c,\lambda}$ is: 
\begin{eqnarray}
J_{\lambda}(\tau_{\lambda},\tau_{c,\lambda},\chi_0) &=&
\sum_{m=1}^M A_{m,\lambda}(\chi_0)\nonumber \\
&&\times \left\{ \exp[-k_{m,\lambda}\,\tau_{\lambda}] +
                                 \exp[-k_{m,\lambda}\,(2\tau_{c,\lambda}-\tau_{\lambda})]\right\}.
\label{jlambda}
\end{eqnarray}
The units of $J_{\lambda}$ are 
erg$\persec\cmtwo\,{\rm sr}^{-1}\,\micron^{-1}$.  
The coefficients, $A_{m,\lambda}(\chi_0)$, are determined using the boundary 
condition at $\tau_{\lambda} = 0$, \ie\ 
$J_{\lambda}(0,\tau_{c,\lambda}, \chi_0) \equiv \chi_0\, J_{{\litl ISRF},\lambda}$,
whence:
\begin{equation}
A_{m,\lambda}(\chi_0) = \frac{\chi_0\,J_{{\litl ISRF},\lambda}}
{\left[ 1 + \exp(-2k_{m,\lambda}\,\tau_{c,\lambda})\right]M}.
\label{coeff}
\end{equation}
The multiplicative factor, $\chi_0$, allows for variations in the nominal
radiation field \citep{mez82,mat83}.  The scattering 
modifiers, \{$k_{m,\lambda}$\}, are the reciprocal 
eigenvalues of an $M\times M$ tridiagonal symmetric matrix, {\bf X}, given 
by Equation (A3) of \citet{fla80}.  If we use the scattering 
phase function introduced by \citet*{hen41}, then the matrix elements are 
functions only of the
dust grain albedo, $\langle \omega_{\lambda}\rangle$, and the mean cosine 
of the 
scattering angle, $\langle g_{\lambda}\rangle$.  The $\langle\rangle$ notation
we use denotes an average of dust optical properties over the grain 
population (see Appendix).  The $l$th off-diagonal 
element of the matrix, $X_{l,l+1}$, equals $l/\sqrt{h_{l-1}h_l}$, where
\begin{equation}
h_l = (2l + 1) ( 1 - \langle \omega_{\lambda}\rangle \langle g_{\lambda}\rangle^l ).
\end{equation}
The matrix is symmetric, so $X_{l+1,l} = X_{l,l+1}$.  In 
practice we truncate the series in Equation \ref{jlambda} to $M = 20$ terms,
since for the wavelengths of interest $k_{21}$ is less than 1\% of the value
of $k_{20}$.  Therefore, {\bf X} is a $20\times 20$ matrix.

Equation \ref{jlambda}
takes as inputs the optical depth, $\tau_{\lambda}$, and the total size of the 
cloud, given by $\tau_{c,\lambda}$, but we can switch our depth scale to 
that of the visual extinction, $\av = 1.086\tau_{\litl v}$.  The visual 
extinction is a convenient depth scale because it can be related linearly 
to the column density of hydrogen nuclei, \nnh\ \citep[eg.,][]{boh78}.  It
can be defined in terms of the wavelength--dependent optical depth
using the extinction curve (see Equation \ref{extincurve}):  
\begin{equation}
\av = 1.086\tau_{\lambda}\,\left(\frac{A_{\lambda}}{\av}\right)^{-1}.  
\label{a_v}
\end{equation}
The central visual extinction, \avc, can be defined similarly in terms 
of  $\tau_{c,\lambda}$.  

We have used Equation \ref{jlambda} to calculate the mean intensity spectrum
inside interstellar clouds (Figure \ref{lambdajlambda}).  In what follows, 
we use the Equation \ref{jlambda} attenuated radiation field to compute
the heating function for dust grains and the photoelectric 
heating function for the gas.  Our assumption of thermal equilibrium 
then allows us to estimate the surface brightness of 100\micron\ and 
60\micron\ radiation from the dust and the [\cii ] emission line
cooling of the gas.

\subsection{Dust and gas heating}

\subsubsection{Thermal equilibrium of dust grains}
The temperature of a spherical grain of radius $a$ and material $i$ (see 
Appendix), measured at depth \av\ in a cloud of central thickness \avc,
 is $T_g(a,i,\av,\avc,\chi_0)$.  It can be determined by equating the power 
absorbed by the grain to the power emitted.  This requires:
\begin{equation}
4\pi^2 a^2 \int_0^{\infty} J_{\lambda} \,Q_{abs,\lambda}(a,i)\, d\lambda
= 4\pi^2 a^2 \int_0^{\infty} B_{\lambda}(T_g) \,Q_{abs,\lambda}(a,i)\, d\lambda,
\label{dustbalance}
\end{equation}
where $B_{\lambda}(T)$ is the Planck function and $Q_{abs,\lambda}(a,i)$ is
the absorption efficiency (see Appendix).  The assumption here is that 
dust grains are in thermal equilibrium.  This
holds for grains with sizes $a\gtrsim 250\AA$, but breaks down for 
smaller particles, which are stochastically 
heated (See \citealt{dra01}, and references therein).  We show in \S5.3.1
that this equilibrium approach does not affect significantly our 
conclusions.

We have solved Equation \ref{dustbalance}
numerically for the function $T_g(a,i,\av,\avc, \chi_0)$.  As an
 example, we plot 
in Figure \ref{heating} the equilibrium temperature of a graphite ($i=1$) 
grain of size $a=0.1\micron$ as a function of \av, for a cloud with 
$\avc = 1.0\,$mag, immersed in a $\chi_0 = 1$ interstellar radiation field.  
At the surface of the cloud, a 0.1\micron\ graphite grain has an equilibrium 
temperature of $T_g \approx 21\,$K.  This is 
comparable to the value independently computed by \citet{li01} for 
``diffuse clouds,'' \ie\ clouds without ISRF attenuation.   For all grain 
sizes $\gtrsim 0.01\micron$ and for both graphite and silicate particles, 
our values of $T_g$ at the surfaces of clouds match those of 
\citeauthor{li01}.  Our model, which computes the grain temperatures
as a function of cloud size, depth, and $\chi_0$, can be used to 
extend the \citeauthor{li01} model to the translucent regime, where
radiative transfer is important.  Figure 
\ref{lambdajlambda} shows how the spectral shape of the radiation field is 
already significantly different at the center of an $\avc =1$ cloud from 
that at the surface.  Obviously for translucent clouds, a depth-dependent 
approach to computing grain temperatures is necessary.

\subsubsection{Photoelectric heating of gas}
The most important way to transfer energy from the 
ISRF to the gas is via the photoelectric effect on dust grains.  
Photoelectric heating of the gas can 
occur when photons with energies larger than the grain work 
function expel electrons.  A typical neutral grain
work function is $\sim 5.5\eV$, so it is mainly FUV photons 
($6\eV < h\nu < 13.6\eV$ ;  $0.2066\micron > \lambda > 0.0912\micron$) that 
contribute to the photoelectric effect.  The photoelectric heating 
mechanism has been explored in detail by \citet{bak94}.  They
found that a fraction, $\epsilon$, of the total FUV energy
absorbed by all grains is available as photoelectron kinetic energy to
heat the gas.  The value of $\epsilon$, the ``photoelectric heating
efficiency,'' is expected to be $\sim 3\%$ for neutral grains \citep{bak94}.  
This can be explained using a simple heuristic argument.  An
average FUV photon has energy of 8\eV, which causes an electron with 2.5\eV\
of kinetic energy to be released from a neutral grain with work function 
5.5\eV.  In other words, 30\% of the incoming photon's energy is liberated.  
The typical photoelectric yield is 10\%, \ie\ 10\% of all photons absorbed
by grains actually ionize them.  Therefore, the fraction of photon energy 
absorbed that gets injected into the gas as kinetic energy is 
$\epsilon \sim 0.1\times 0.3 = 0.03$.  

The neutral grain photoelectric efficiency is an upper limit to the
actual value of $\epsilon$.  For ionized grains, the work 
functions are higher than 5.5\eV, and $\epsilon$ is less than the
neutral grain limit.  In the
cold neutral medium, less than 50\% of grains with 
$a\lesssim 15\AA$ are expected to be ionized \citep{li01,wei01}, implying
relatively high photoelectric efficiency in translucent clouds.

We used our cloud model to estimate the gas heating in translucent clouds.  To
compute the total amount of FUV energy absorbed by grains we first estimated
the mean absorption cross section per H~atom as a function of wavelength, 
$\langle\sigma_{\lambda}\rangle$:
\begin{equation}
\langle\sigma_{\lambda}\rangle = 
\sum_{i=1}^{2} \int_{3.2\AA}^{a_{\litl max}}\pi a^2\, C_i \, 
Q_{abs,\lambda}(a,i)\, a^{-3.5}\, da,
\label{sigma}
\end{equation}
where $\{C_i\}$ and $a_{\litl max}$ are defined in the Appendix.  
\citet{bak94} estimated that
approximately one half of the heating in interstellar clouds is caused by 
tiny carbon grains and Polycyclic Aromatic Hydrocarbons (PAHs)
with sizes less than 15\AA, and the other half comes from
larger grains with $15\AA < a < 0.01\micron$.  The size
distribution in the \citet{mat77} model (Equation \ref{graindist}) has 
a minimum cutoff, $a_{\litl min} = 0.005\micron$, so we had to extend the 
grain parameter calculations of \citet{dra84} to include smaller graphite 
grains.  Following \citet{bak94}, we assumed that for grains with $a \ll 
\lambda$ (the Rayleigh limit) the absorption and scattering efficiencies
for graphite grains,
$Q_{abs,\lambda}(a,1)$ and $Q_{sca,\lambda}(a,1)$, respectively, scale with 
grain volume.  We also assumed that the mean scattering
cosine for graphite, $g_{\lambda}(a,1)$, follows the Mie expansion given in 
\citet[their Equation 8]{lao93}.  

The local gas heating function, $\Gamma$, is derived by integrating the 
mean intensity of the
local radiation field, weighted by 
the absorption cross section per H~atom and the photoelectric efficiency,
over all FUV wavelengths and solid angles:
\begin{equation}
\Gamma= 4\pi\epsilon\,\int_{0.0912\micron}^{0.2066\micron}
J_{\lambda} \, \langle\sigma_{\lambda}\rangle \, d\lambda.
\label{heating_eq}
\end{equation}
The units of $\Gamma$ are erg$\,\persec\,({\rm H~atom})^{-1}$.  In Figure
\ref{heating} we show the behavior of $\Gamma$ as a function of \av\
for a cloud with $\avc = 1.0\,$mag, $\epsilon = 0.043$, and $\chi_0=1$.  Near 
the surface of the cloud ($\av\sim 0.3\,$mag), the heating from our model
equals the average cooling observed with {\sl COBE} for high--latitude
\hi\ gas, $2.65\times\expo{-26}\,{\rm erg}\persec\,({\rm H~atom})^{-1}$
\citep{ben94}.

\subsection{Dust and gas cooling}

The grain temperature and the gas heating function allow us to
derive the $FIR$ intensity and the [\cii ] intensity, respectively, 
towards translucent clouds
to compare with observations of these important cooling sources.  In
the context of our proposed model these are the {\it sole} mechanisms of
cooling.

\subsubsection{FIR Emission}
Integrating over the grain distribution, the local cooling function of 
dust by thermal emission is:
\begin{equation}
\Lambda_{{\litl d},\lambda} = \sum_{i=1}^{2}
\int_{a_{\litl min}}^{a_{\litl max}}
 B_{\nu}(T_g) \,\pi a^2\, Q_{abs,\lambda}(a,i)\, a^{-3.5}\,da ,
\label{dustcool}
\end{equation}
where
\begin{equation}
\frac{B_{\nu}(T)}{\mjysr} = \frac{3.97\times\expo{-11}}
                 {{\rm exp}(1.44\times\expo{-4}\micron\cdot\K/\lambda\, T) - 1}
\label{planck}
\end{equation}
is the Planck function.  Since $B_{\nu}[T]$ is
in \mjysr, the units of $\Lambda_{\litl d}$ are 
$\mjysr\cdot\cmsquared\,({\rm H~atom})^{-1}$.  This is equivalent to \mjysr\
per H column density, \nnh.  The emergent thermal dust 
emission spectrum can be derived by converting to \mjysr\ per \av\,
using the standard relation, $\nnh \,(\cmtwo) = 1.87\times\expo{21}\av\,
({\rm mag})$, integrating $\Lambda_{\litl d}$ from the surface of the cloud to
the center, and multiplying by 2:
\begin{equation}
I_{\lambda}(\avc,\chi_0) = 3.74\times\expo{21} \int_{0}^{\avc} 
\Lambda_{{\litl d},\lambda}(\av,\avc,\chi_0)\, d\av .
\label{firspec_eq}
\end{equation}
The factor 2 is necessary because the integral is from the edge to the
center of a symmetrical one-dimensional cloud.  

\subsubsection{FIR Spectra}
Model FIR spectra, $I_{\lambda}(\avc,\chi_0)$, are
plotted in Figure \ref{firspec} for $\chi_0 = 1$.  
To check our model for internal consistency, we realize that the 
integrated FIR emission should be less than or equal to the incident ISRF
integrated intensity, $2.76\times\expo{-2}\,
{\rm erg}\cmtwo\persec$ (Equation \ref{isrf_integrated}).  For 
$\avc = (0.5,1.0,2.0,5.0)$,
the integrated FIR flux from the clouds is $(1.00,1.53,2.03,2.64)\times
\expo{-2}\,{\rm erg}\cmtwo\persec$.  Thus, only for extremely opaque 
clouds ($2\avc \sim 10.0\,$mag) is most of the input radiation recovered
as FIR emission, and the input flux is never exceeded.

We can use Figure \ref{firspec} to predict that clouds with large $\avc$ 
appear ``colder'' than clouds with small $\avc$, provided $\chi_0$ 
is held constant.  The long wavelength spectrum mimics
 the emission by dust at a single temperature.  Thick gray curves in 
Figure \ref{firspec} show modified blackbody spectra for 
clouds with the same sizes (values of \avc) as the model clouds, but with
different ``dust temperatures,'' $T_d = $(19,18,16.9,15.1)\,K, for $\avc = 
(0.5,1.0,2.0,5.0)$.  Here the blackbody function at $T_d$ was 
modified by the dust emissivity, assumed to be proportional to 
$\lambda^{-2}$.  The modified blackbody curves reproduce 
the model spectra quite well for wavelengths longward of the peak.  

We find that a thicker cloud produces more 
emission than a thinner cloud, but its spectrum 
peaks at a longer wavelength (vertical arrows), yielding a lower 
``dust temperature,'' as shown by the gray curves.  The same behavior was 
noticed for
a large sample of high--latitude clouds by \citet{rea98}, who found
that the average color temperature for dust mixed with
HLC atomic gas is $(17.0\pm 0.3)\K$, whereas for dust mixed with HLC 
molecular gas the average color temperature is 
$(14.7\pm 0.2)\K$.  This happens because, even though opaque clouds 
absorb more of the ISRF than transparent clouds, most of the dust 
in their interiors is exposed to an attenuated version of the ISRF 
that is deficient in high energy photons.  

Our model can be used to estimate the 60 and 100\micron\ 
surface brightness, \irasc\ and \irasd, towards translucent clouds, as
well as the {\sl IRAS}--derived $FIR$ integrated surface brightness 
(Equation \ref{fir_def}).  Figure \ref{cooling} plots
\irasc, \irasd, and $FIR$ versus \avc\ for $\chi_0 = 1$.  Note that even
though $FIR$ increases with increasing \avc, the rise is not linear, \ie\
the $FIR$ {\it emissivity} ($FIR$/\nnh) of translucent clouds 
decreases with increasing \nnh, as is observed \citep[see][]{rea98}.

Another way to demonstrate that thicker clouds have ``cooler'' dust 
emission is to compare the surface brightness ratio \irasc/\irasd\
with visual extinction, as we have done empirically for MBM-12.  Gray 
curves on Figure
\ref{mbm12_av_color} show that, in our model, the 60/100 color decreases 
with \av\ (= 2\avc) and increases with $\chi_0$.  The model agrees 
remarkably well with the MBM-12 data for $\chi_0 = 1.6\pm 0.2$.  Such
behavior was already noticed and discussed qualitatively for 
translucent clouds in the Lynds 134
cloud complex (which contains two objects in our sample, MBM-37 and 
MBM-33) \citep{lau91,lau95}.   Cuts across the densest regions of L134
\citep{lau95} show that the 60/100 color decreases as B-band extinction, 
$A_B$, increases.  We have compared the \citeauthor{lau95} L134
data to our model (using $1.4 A_B \approx \av$), and we find a good match for
$\chi_0 = 1.4-1.6$ and $\av = 0-5$.  There is a sudden drop in 60/100 
for extremely opaque ($\av > 5$) lines of sight towards the L134 complex 
\citep{lau91,lau95}, which the model does not predict.  The authors 
attribute this to a change in dust properties (see \S5.3).

\subsubsection{[\cii ] emission}

We estimate the [\cii ] integrated intensity towards translucent clouds.  
Recall our key assumptions that the gas cools only through emission in the 
(\ppp ) transition
of \cii, and that the gas is in thermal equilibrium.  We set the cooling, 
$\Lambda$ (\ie\ the [\cii ] emission per \nnh), equal to the gas heating, 
$\Gamma$, given by Equation \ref{heating_eq}.  The integrated intensity of
the [\cii ] emission is:
\begin{equation}
\icii(\avc,\chi_0,\epsilon) = \frac{3.74\times\expo{21}}{4\pi}\, \,
\int_{0}^{\avc} \Gamma(\av,\avc,\chi_0,\epsilon)\, d\av .
\label{ciicool}
\end{equation}

We plot in Figure \ref{cooling} the [\cii ] integrated intensity, \icii, as a 
function of \avc\ for clouds with $\chi_0 = 1$ and $\epsilon = 0.043$.  
The two upper curves highlight the differences 
between gas and dust heating by interstellar radiation.  The qualitative 
behavior of both curves is the same.  That is, both $FIR$ and \icii\
rise almost linearly, implying linear increases in column-averaged dust 
and gas
heating with extinction; then both curves show a decrease in slope, implying 
that clouds with central
extinction larger than a transition size produce much less emission at 
depths greater than this size.  The effect of reaching the transition sizes
is different for the two tracers.  The \icii\ emission becomes
flat at $\avc\sim 0.5\,$mag, whereas $FIR$ only shows a decrease in
slope, without flattening even for $\avc= 2.0\,$mag.  In the interior 
of a translucent cloud the mean integrated intensity of FUV 
radiation is much smaller than that of optical radiation 
(refer to Figure \ref{lambdajlambda}).  Dust grains are heated by ISRF 
photons of all wavelengths, but photoelectrons, which lead to 
[\cii ] emission, are ejected only by FUV photons.  Thus, one 
expects the [\cii ] emission to extinguish before the $FIR$ emission does.

\subsubsection{Linear and Flat Behavior}

The model for $FIR$ and [\cii ] emission depends on three parameters 
(see Equations \ref{firspec_eq} and 
\ref{ciicool}):  (1) the cloud column density, given by \avc;
(2) the incident ISRF flux, $\chi_0$, in units of the expected flux at 
the Sun; and (3) the photoelectric heating efficiency, $\epsilon$.  We have
computed cloud models for \avc\ ranging from 0 to 5\,mag, and $\chi_0$ 
ranging from 0.2 to 4.0.  The
[\cii ] results scale linearly with the photoelectric heating 
efficiency (Equation \ref{heating_eq}), so it was trivial to 
vary $\epsilon$.  

Figure \ref{cp_fir_model} shows plots of \icii\ as a function of $FIR$ for 
four different combinations of $\chi_0$ and $\epsilon$.  The curves show two 
types of behavior:  ``linear'' behavior for low values
of $FIR$; and ``flat'' behavior for high values of $FIR$.  This is due
to the difference between dust and gas heating mentioned above.  The linear
behavior occurs when \avc\ is low enough that both dust and gas heating 
increase linearly with \avc.  The flat behavior occurs when \avc\
is high enough for FUV heating of the gas to become negligible in cloud 
interiors, causing [\cii ] emission to extinguish, but not high enough for 
optical heating of dust grains to become negligible, so FIR
emission is still produced.  

We have analyzed models for a range of 
$\chi_0$ and $\epsilon$ values.  The behavior of \icii\ and $FIR$
in the linear regime can be approximated by:
\begin{equation}
{\rm I_{\litl CII}^{\litl LINEAR}} = FIR\times
(0.5634\pm 0.0003)\,\chi_0^{-(0.267\pm 0.001)}\,\epsilon .
\label{linear_eq}
\end{equation}
Errors represent formal uncertainties in fits to the grid of cloud
models.  The slope of the linear behavior (Equation \ref{linear_eq}) depends 
primarily on $\epsilon$ and only weakly on $\chi_0$.  The slope should
not depend at all on $\chi_0$, since both \icii\ and the true FIR integrated 
intensity should be linearly proportional to $\chi_0$.  Recall, however, that 
for comparison with observations we have approximated $FIR$ using \irasc\ and 
\irasd\ (see Equation \ref{fir_def}), \ie\ we bias the $FIR$ integral 
artificially towards short far-infrared wavelengths.   As $\chi_0$ 
increases, the 
short wavelength surface brightness increases faster than linearly, 
since the peak of the FIR spectrum shifts to shorter wavelengths.  Thus our 
$FIR$ values should also increase with $\chi_0$ faster than linearly,
while the integrated spectrum increases linearly.

The [\cii ] intensity in the flat regime can be approximated by:
\begin{equation}
\frac{\rm I_{\litl CII}^{\litl FLAT}}{\intunit{-6}} = (149.05\pm 0.07)
\,\chi_0\,\epsilon .
\label{flat_eq}
\end{equation}
The flatness occurs because
at high opacity all incident FUV radiation has been absorbed
and converted to photoelectric heating (with efficiency $\epsilon$).  
The maximum possible [\cii ] intensity ought to equal the total integrated 
FUV intensity in the ISRF, multiplied by $\epsilon$:
\begin{equation}
\chi_0\,\epsilon\,\int_{0.0912\micron}^{0.2066\micron} J_{{\litl ISRF},\lambda} d\lambda = 
1.44\times\expo{-4}\,\chi_0\,\epsilon.
\label{fuv_integrated}
\end{equation}
Equations \ref{fuv_integrated} and \ref{flat_eq} are indistinguishable to
within 4\%.  We superimpose gray line segments on the curves drawn in Figure 
\ref{cp_fir_model}, representing the linear and flat portions of the curves, 
computed using the approximations in Equations \ref{linear_eq} and 
\ref{flat_eq}.  

We define the transition value of $FIR$ to be the value 
when 
${\rm I_{\litl CII}^{\litl LINEAR}} = {\rm I_{\litl CII}^{\litl FLAT}}$, 
\ie\
\begin{equation}
\frac{(FIR)^{\litl TRANSITION}}{\intunit{-6}} = (264.55\pm 0.19)\,\chi_0^{(1.267\pm 0.001)}.
\label{transition_eq}
\end{equation}
The $FIR$ value at which a transition between the linear and flat
regimes occurs depends wholly on $\chi_0$.  We can use 
Equation \ref{fir_def} to derive a 100\micron\ surface brightness 
at the linear to flat transition for typical CNM gas ($\irasc = 0.16\irasd$):
\begin{equation}
\frac{{\rm I}_{\litl 100}^{\litl TRANSITION}}{\mjysr} = (14.86\pm 0.01)\,\chi_0^{(1.267\pm 0.001)}.
\label{transition_irasd}
\end{equation}

\subsection{Statistical Comparison of Data with Observations}

\begin{figure}
\plotone{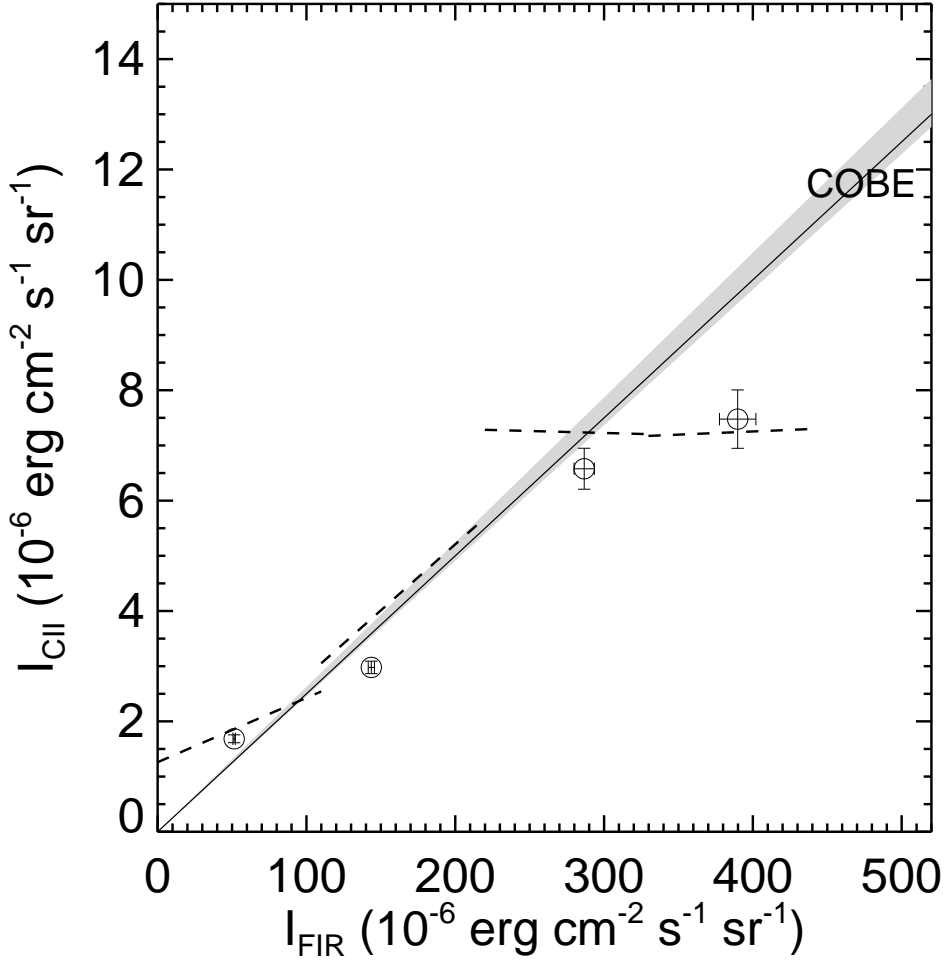}
\caption{The [\cii ] emissivity of HLCs decreases as $FIR$ increases.  
Dashed line segments display weighted linear fits to the HLC data 
(Figure \ref{cp_fir}), separated into bins of width 
$\expo{-4}\erg\cmtwo\persec\,{\rm sr}^{-1}$ along the $FIR$-axis.
Circles with error bars denote the weighted mean values of $\icii$ and $FIR$
in these bins.  The gray wedge represents the 
range of {\sl COBE} data for average Milky Way sources, as described in the 
caption to Figure \ref{cp_fir}.  The solid line is the fit to our HLC data 
(Figure \ref{cp_fir}). 
\label{cp_fir_bins}}  
\end{figure}

\begin{figure}
\plotone{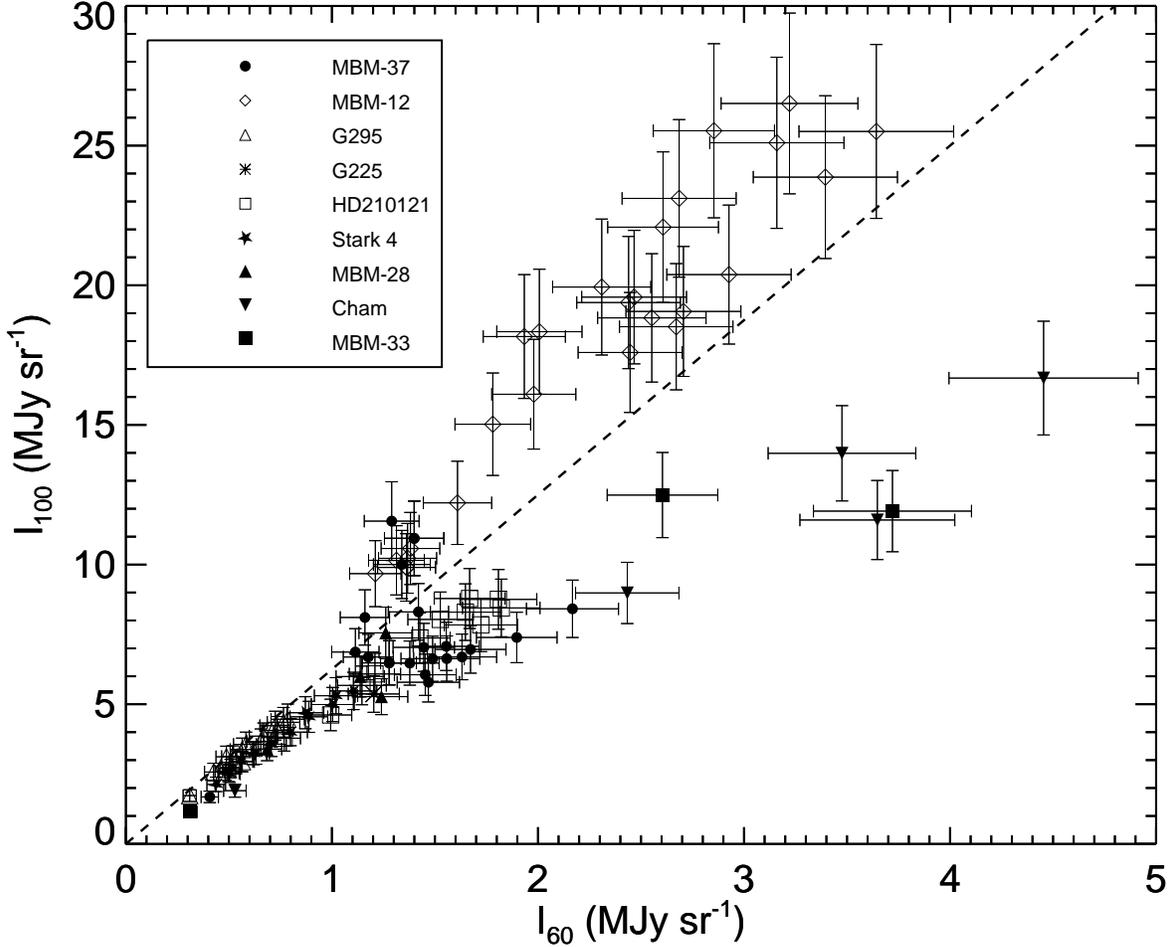}
\caption{Surface brightness at 100\micron\ versus that at 60\micron\ for our 
HLC sample.  A 
dashed line indicates the mean behavior derived using {\sl COBE} DIRBE
observations of Galactic gas associated with \hi ,
$\irasc = 0.16\,\irasd$ \citep{dwe97}.  This line is not a fit to the 
data.  We identify two populations based on the 60/100 color:  a ``warm''
population for which $\irasc\gtrsim 0.16\irasd$; and a ``cold'' population 
for which $\irasc < 0.16\irasd$.
\label{iras_100_60}}
\end{figure}

\begin{figure}
\plotone{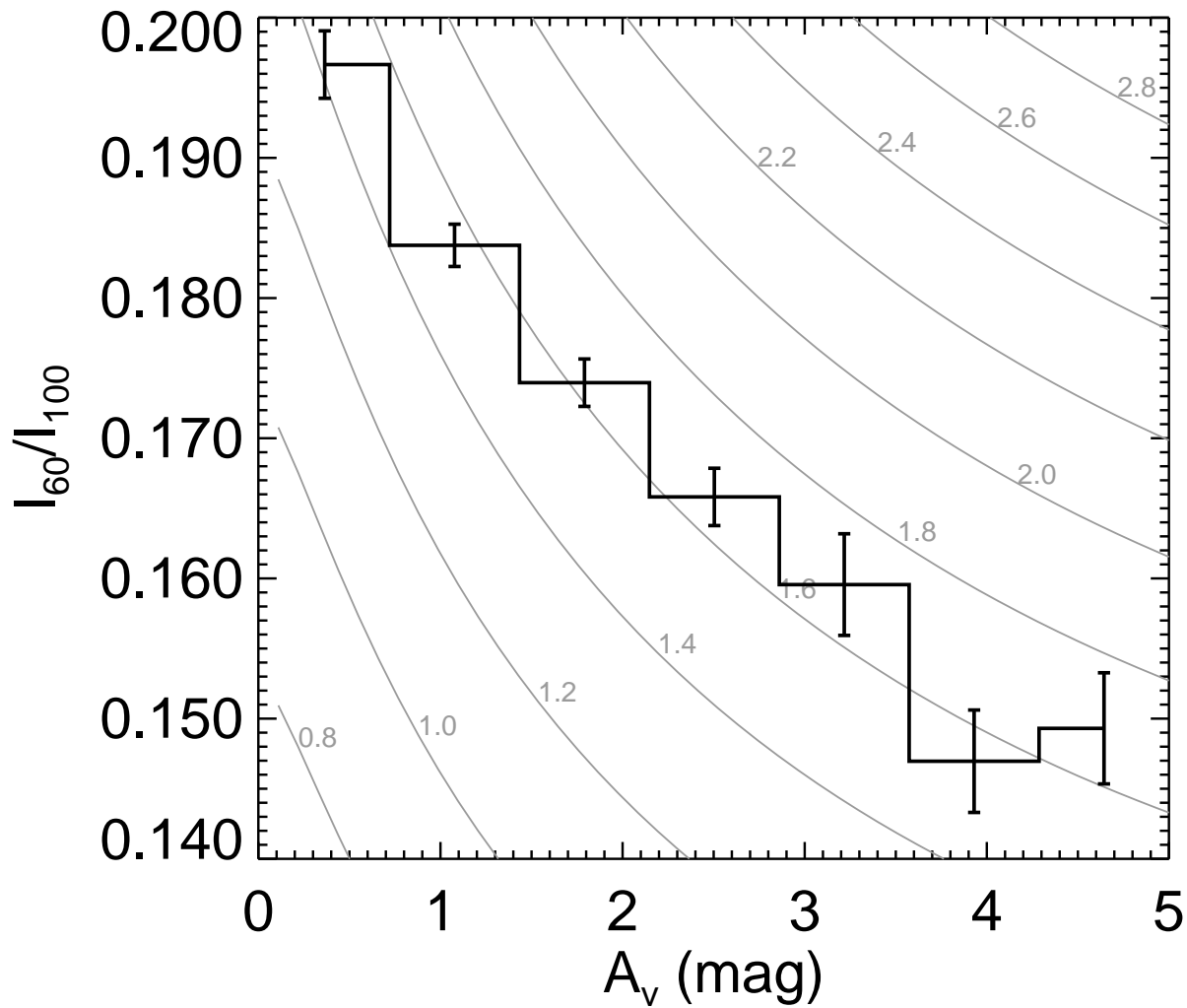}
\caption{The surface brightness ratio, \irasc/\irasd\, decreases as
a function of visual extinction for cloud MBM--12.  Image data for over
2,000 pixels are averaged in bins of width $\Delta\av = 0.71\,$mag.  Note
that a small adjustment in 60/100 represents a large change in \av.  Models 
of the thermal emission from dust grains in clouds immersed in the
interstellar radiation field (see \S 4.3.3) are represented as gray 
curves.  The contours are labeled with $\chi_0$, the total ISRF flux measured 
in units of the local value.  
\label{mbm12_av_color}}
\end{figure}

\begin{figure}
\plotone{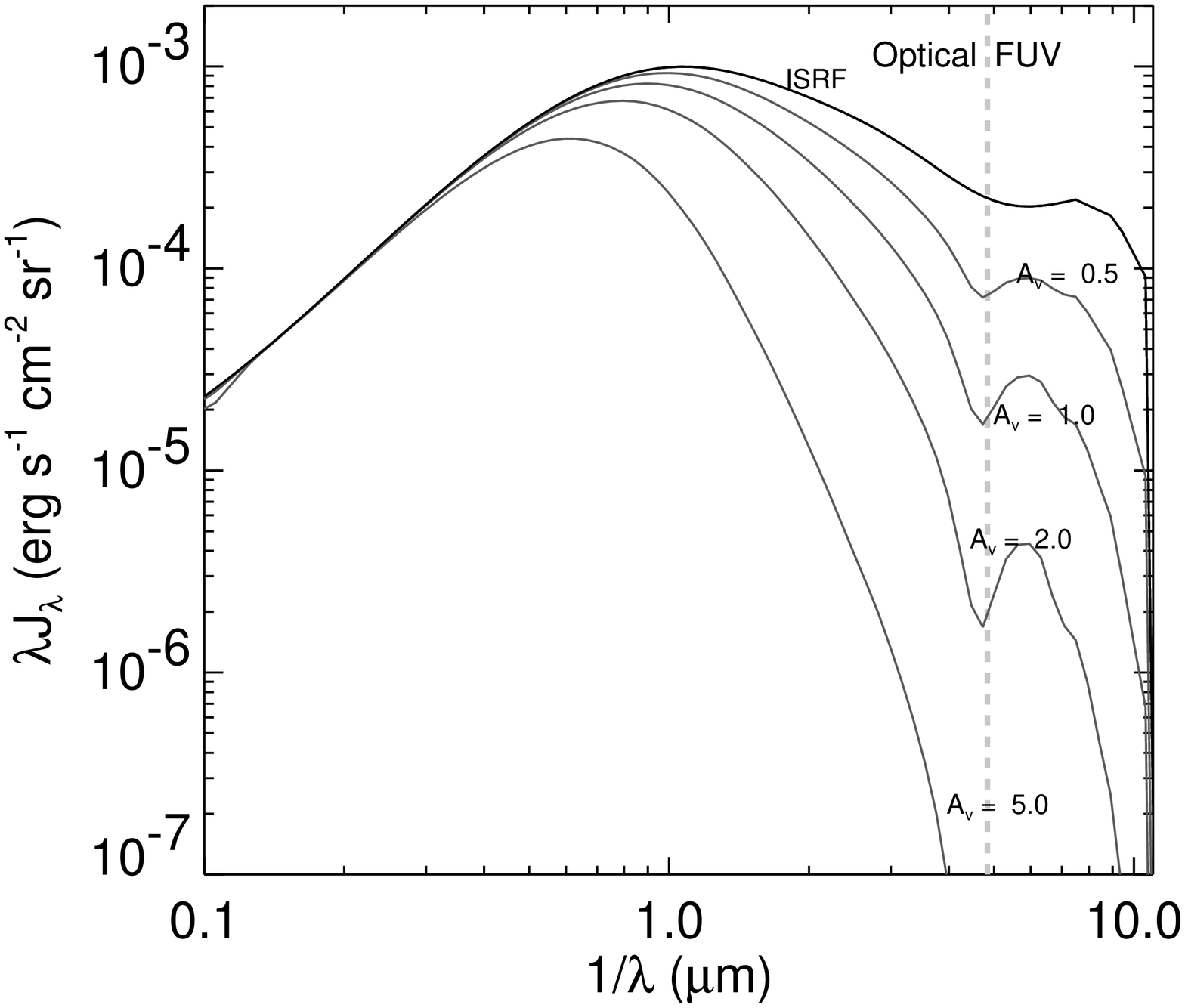}
\caption{The interstellar radiation field (ISRF) and the attenuated spectrum
for four model clouds.  The top curve shows the ISRF spectrum as given
by \citet{mez82} and \citet{mat83}, plus the 2.7\K\ cosmic
background.  This curve defines $\chi_0 = 1$.  The lower curves plot the
wavelength-weighted mean intensity of the radiation field
at the centers of model clouds with central visual extinction, $\avc$,
equal to 0.5, 1.0, 2.0, and 5.0 magnitudes.  A dashed vertical line at
$\lambda = 0.2066\micron$, or photon energies of 6\eV, denotes the
optical/far-ultraviolet (FUV) transition.  Photoelectric heating occurs
to the right of this line.
\label{lambdajlambda}}
\end{figure}

\begin{figure}
\plotone{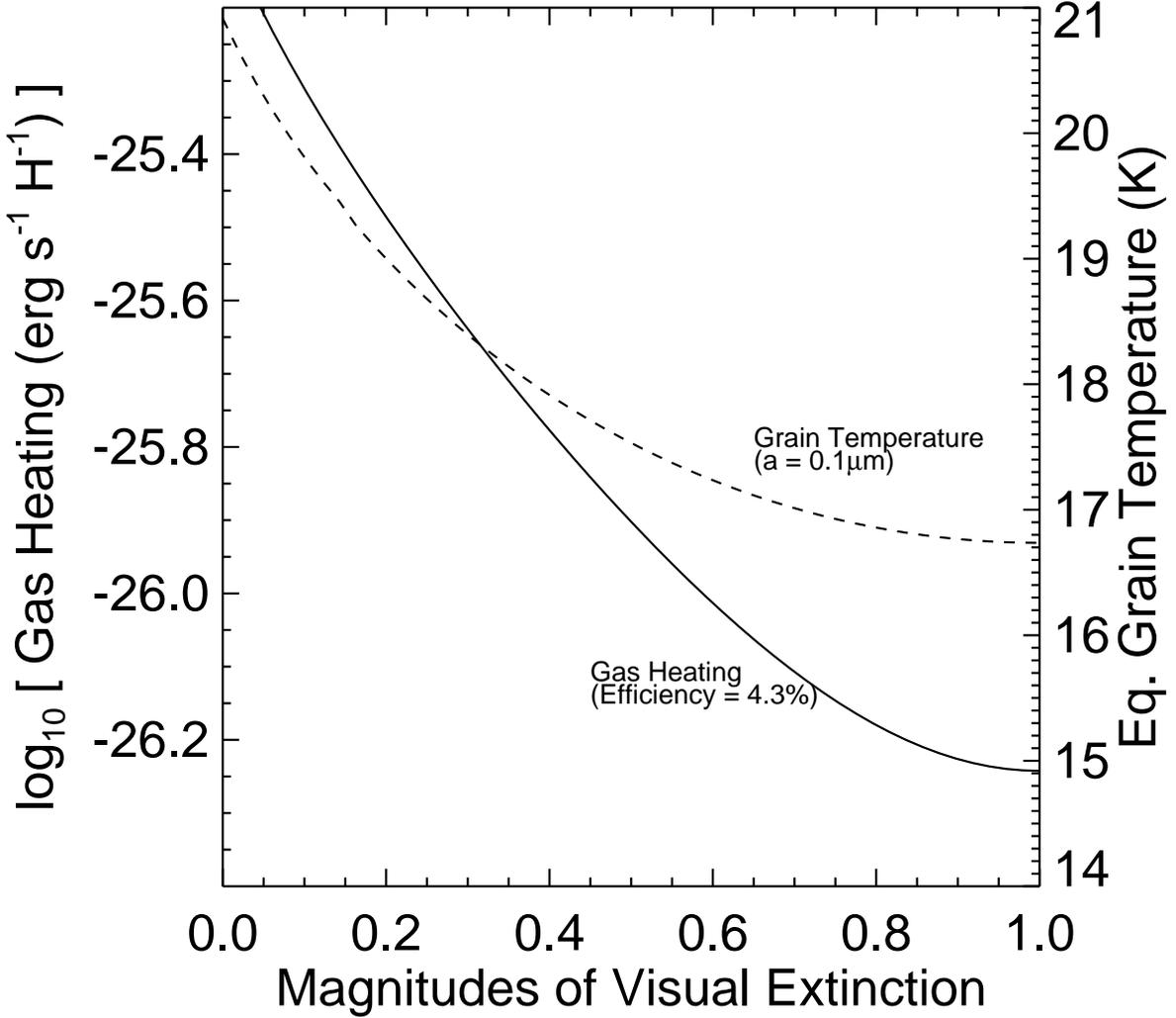}
\caption{The heating of an interstellar cloud with 
$\avc = 1.0\,$mag.  The dashed line (right scale) shows
the equilibrium temperature of graphite grains of size $a=0.1\micron$,
plotted as a 
function of \av.  The solid line (left scale) shows the gas heating
function in the same cloud, assuming the photoelectric efficiency is 4.3\%.  We
only show the profiles for half of the cloud, since the profiles are 
symmetric about $\av = \avc$.
\label{heating}}
\end{figure}

\begin{figure}
\plotone{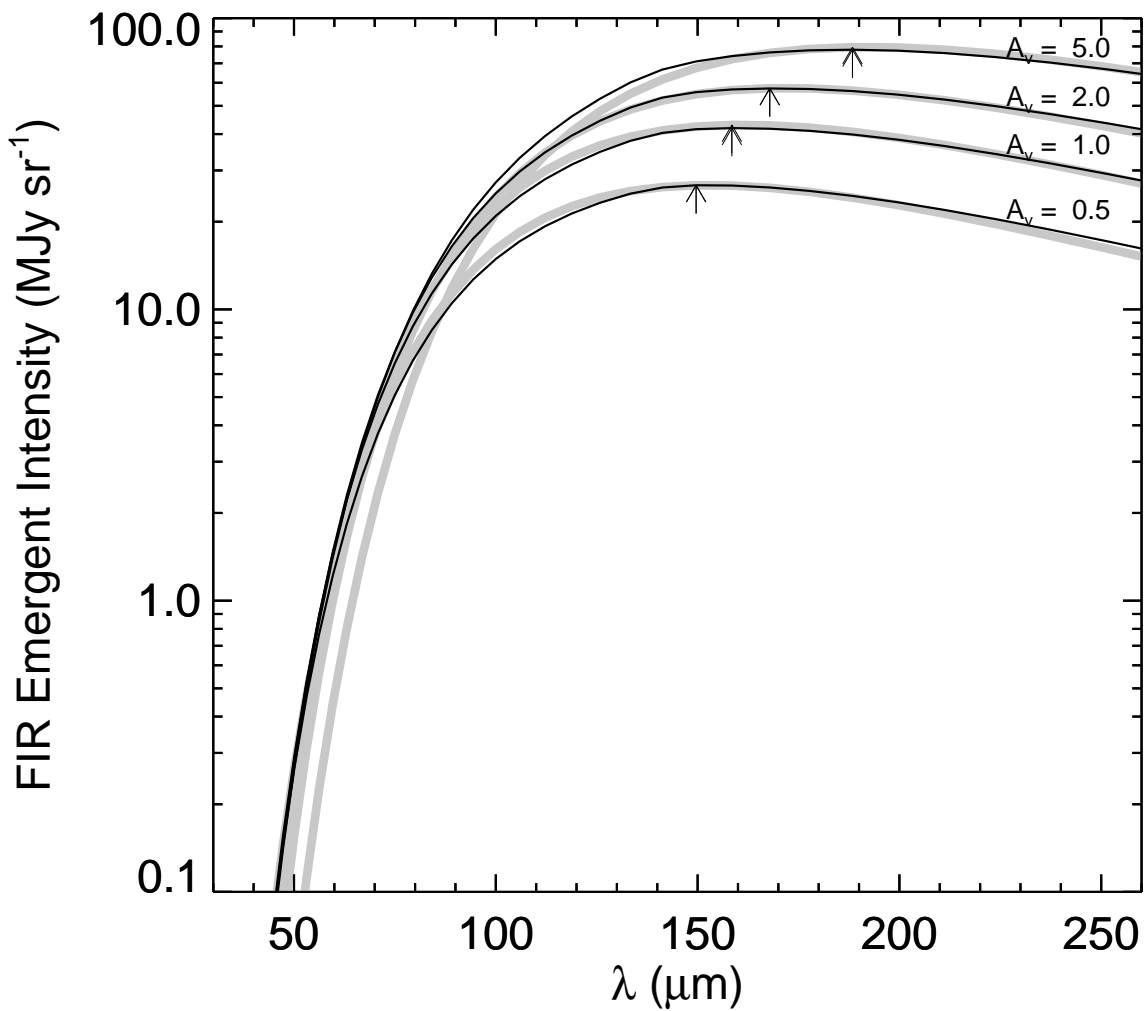}
\caption{The far-infrared (FIR) surface brightness spectra for thermal
equilibrium dust emission from model 
clouds.  Shown here are the spectra for 
clouds with $\chi_0 =1$ and $\avc = 0.5$, 1.0, 2.0, 
and 5.0 magnitudes.  Vertical arrows indicate the peak of 
each curve, demonstrating that more optically thick clouds appear ``colder.''
  Thick gray curves show modified blackbody emission for clouds with the same 
values of \avc, but with constant ``dust temperature,'' $T_d = $19, 18, 16.9,
and 15.1\,K, respectively.  Dust emissivity is assumed to vary as 
$\lambda^{-2}$.
\label{firspec}}
\end{figure}

\begin{figure}
\plotone{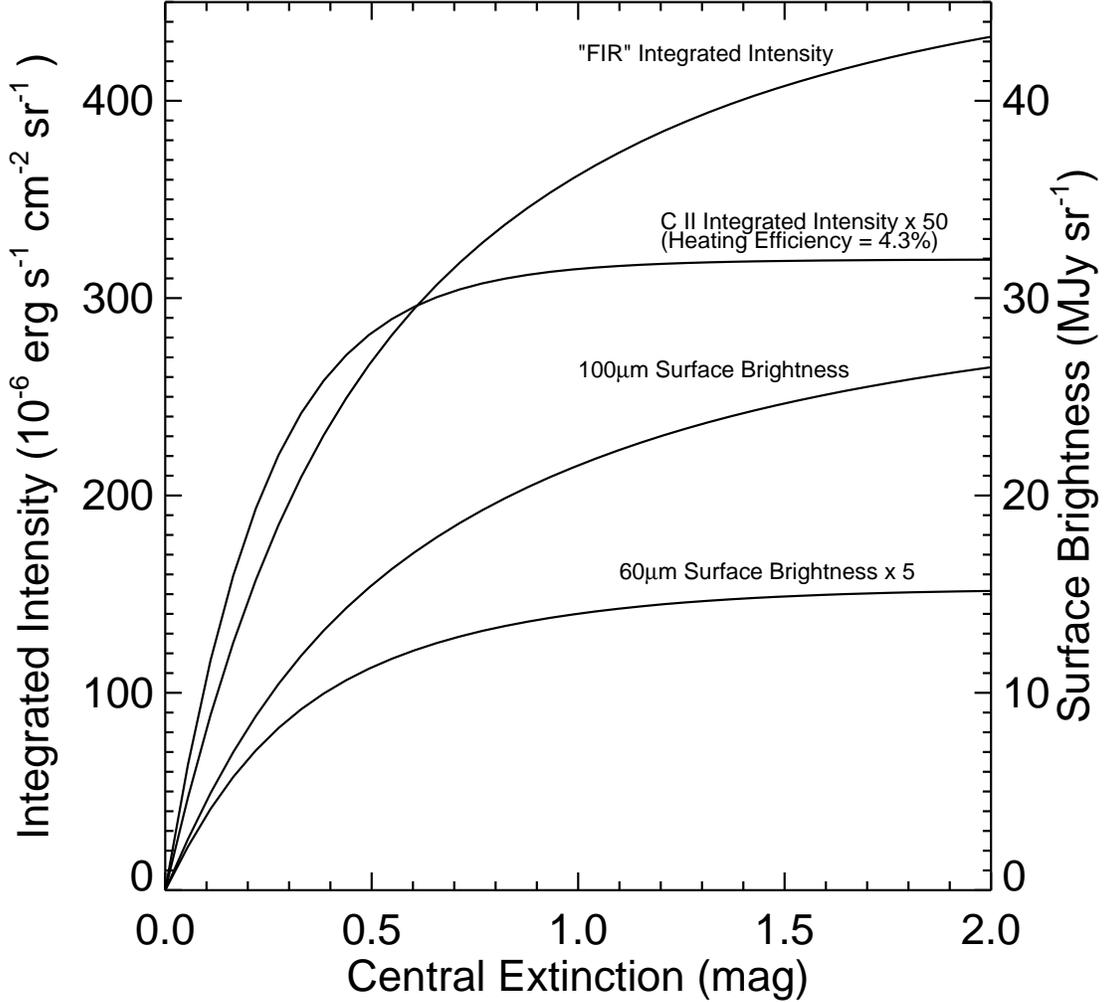}
\caption{The dust and gas cooling towards translucent clouds with $\chi_0 = 1$
and $\epsilon = 4.3\%$.  The two lower curves (right scale) represent 
emission in the 
60\micron\ and 100\micron\ {\sl IRAS} bands from thermal dust grains, 
``observed''
toward a set of model clouds with different column density, given by
\avc.  The \irasc\ values have been multiplied
by 5 for clarity.  The two upper curves (left scale) represent the ``FIR'' 
integrated intensity, derived using \irasc\ and \irasd\ 
(Equation \ref{fir_def}); and the [\cii ] integrated intensity, multiplied 
by 50 for clarity.  
\label{cooling}}
\end{figure}

\begin{figure}
\plotone{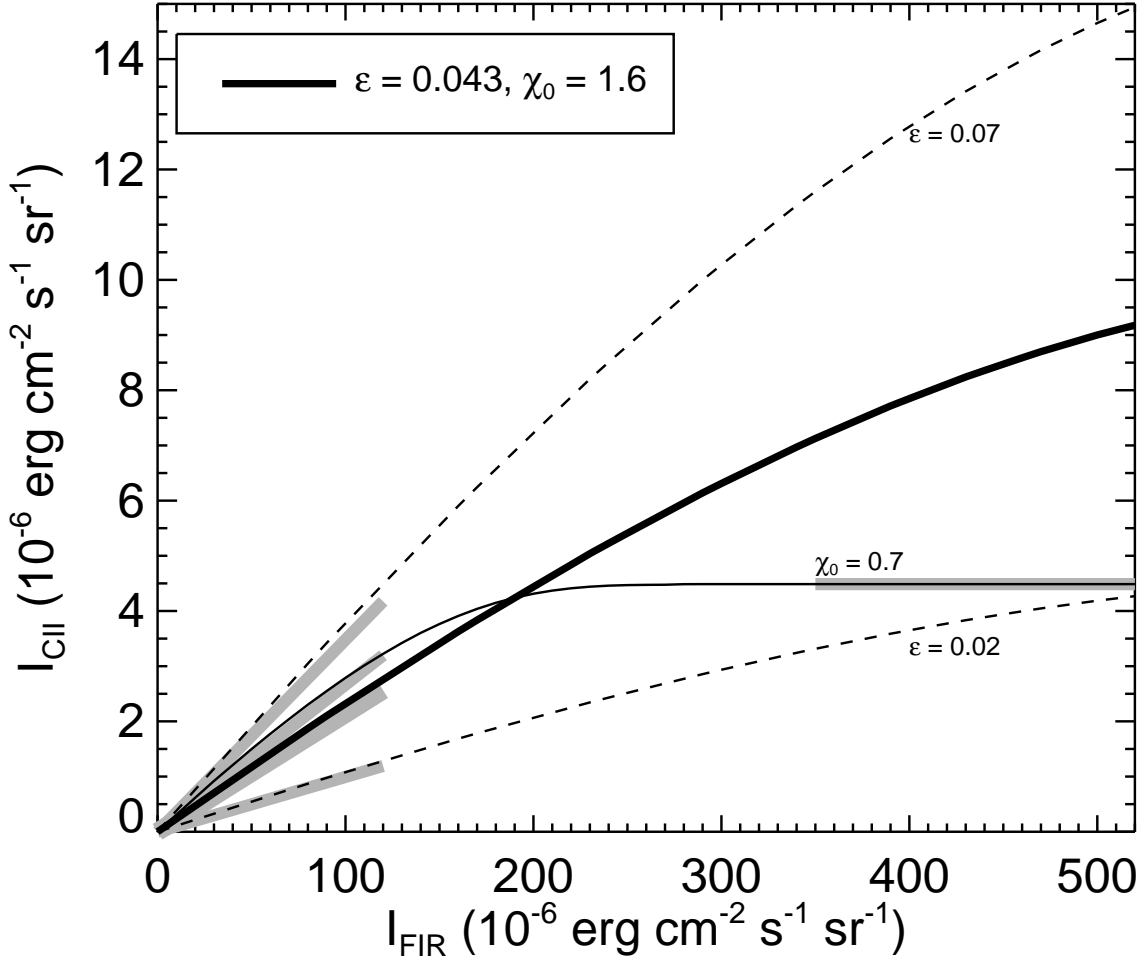}
\caption{Observables \icii\ and $FIR$ for model translucent 
clouds.  Each curve shows the behavior as we vary \avc, for 
four combinations of the surface radiation field, $\chi_0$, and 
the photoelectric heating efficiency, $\epsilon$.  On the \icii --$FIR$
plane, \avc\ increases from left to right and from bottom to top.  The 
thick solid curve shows the best fit HLC values (see
Figure \ref{cp_fir_chi2}), $\chi_0=1.6$ and $\epsilon=4.3\%$.  The
thin solid curve shows the effect of decreasing the ISRF intensity to
$\chi_0 = 0.7$, keeping $\epsilon$ fixed at 4.3\%.  Dashed curves show 
the effect of changing the heating efficiency
from $\epsilon = 2\%$ to $7\%$, holding $\chi_0$ steady at 1.6.  
Gray line segments represent the linear portions of the 
curves and the flat portion of the $\chi_0 = 0.7$ curve, calculated according
to the approximations in the text (Equations \ref{linear_eq} and \ref{flat_eq}).  
\label{cp_fir_model}}
\end{figure}

We can constrain $\epsilon$ and $\chi_0$ by comparing our model for the 
[\cii ] and $FIR$ emission of
translucent clouds to the {\sl ISO} observations of high--latitude clouds.  
First we use the simple equations
that we derived for the linear, flat, and transition zones.  Recall that the 
observations can be fit by a straight line with 
slope $2.5\times\expo{-2}$ (Equation \ref{cp_fir_fit}).  This implies 
that many of the positions observed are in the linear regime.  Substitution 
of the slope into Equation
\ref{linear_eq} gives 
$\chi_0^{-0.267}\epsilon = 0.044$.  The binned
data, however (Figure \ref{cp_fir_bins}), show a statistically significant 
departure from this slope at around $FIR = 3\intint{-4}$.  This appears to
be the transition between the linear and flat regimes.  
Equation \ref{transition_eq} gives $\chi_0 \approx 1.1$, thus 
$\epsilon \approx 0.045$. 
 
A more 
sophisticated way to determine $\epsilon$ and $\chi_0$ for translucent clouds
is to compare a grid of models to the data, and find the best
fit.  Recall that cloud column density, or central visual extinction, is the
independent variable, whereas $FIR$ and \icii\ are 
predicted by the model.  Given a grid of models, we can ``work backwards''
from the observed $FIR$ surface brightness to determine \avc\ as a function 
of $\epsilon$ and $\chi_0$.  Effectively $FIR$ becomes the independent 
variable.  The resulting \avc\ numbers
yield a set of predicted [\cii ] integrated intensities,
${\rm I_{{\litl CII},i}^{\litl PREDICTED}}$ that we can compare with
the data.  We evaluate
the ``goodness of fit'' as a function of $\epsilon$ and $\chi_0$:
\begin{equation}
L^2\,(\epsilon,\chi_0) \equiv \sum_i\frac 
{\left(\rm I_{\litl CII}^{\litl PREDICTED}-{\rm I_{{\litl CII},i}} \right)^2}
{(\sigma_{\litl CII}^{\litl PREDICTED})^2 + \sigma_{{\litl CII},i}^2},
\label{goodness_eq}
\end{equation}
where the sum is over all data points.  
This is the standard $\chi^2$ parameter---we label it $L^2$ here
so as not to confuse the reader with the ISRF intensity, $\chi_0$.  Here 
$\sigma_{\litl CII}^{\litl PREDICTED}$ is computed the same way as
${\rm I_{{\litl CII},i}^{\litl PREDICTED}}$, using the $FIR$ observed 
uncertainty range, $FIR_i\pm\sigma_{FIR,i}$.

\begin{figure}
\plotone{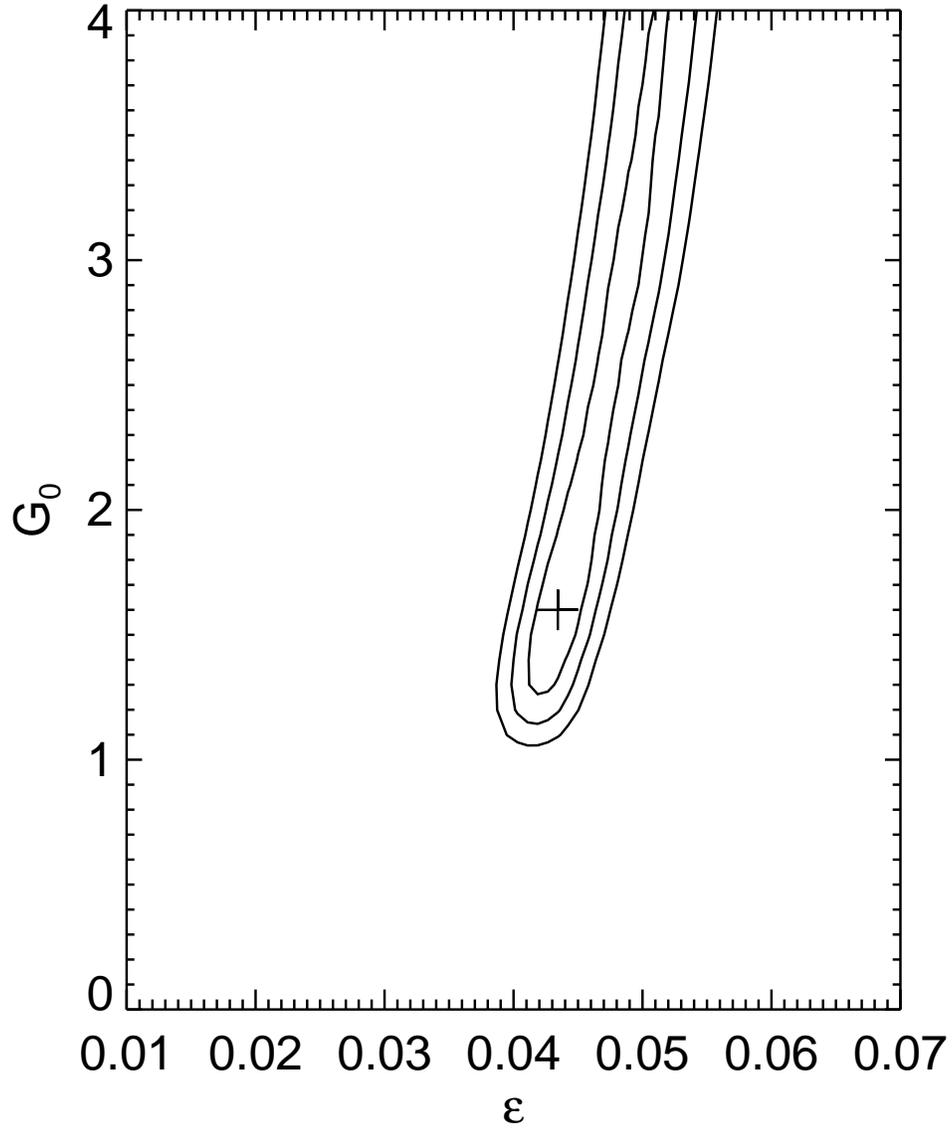}
\caption{Statistical comparison of observed [\cii ] integrated intensity
and $FIR$ surface brightness with our model, for a grid of $\epsilon$
and $\chi_0$ values.  Contours (smallest to largest area) show the 1,
2, and 3$\sigma$ confidence intervals in the fit of the model to
the data shown in Figure \ref{cp_fir}.  The cross represents the best match, 
$\epsilon = 0.043$ and $\chi_0 = 1.6$.  
\label{cp_fir_chi2}}
\end{figure}

Figure \ref{cp_fir_chi2} shows contours of constant $L^2$ in the  
($\epsilon$,$\chi_0$) plane, representing 68.3\%, 95.4\%, and 99.7\%
confidence in the fit of our model to the data.  These are equivalent to
the 1, 2, and 3$\sigma$ confidence intervals for Gaussian statistics.  The 
minimum $L^2$ point, indicated 
by a cross in Figure \ref{cp_fir_chi2}, gives the most probable values,
$\epsilon = 0.043{{+0.010}\atop{-0.002}}$ and $\chi_0 = 1.6$.  We 
derive a lower ($-3\sigma$) limit of $\chi_0 \ge 1$, but cannot derive
an upper limit because the Figure \ref{cp_fir_chi2} contours of constant 
$L^2$ continue indefinitely in the $\chi_0$ dimension.  Apparently a 
complete transition to flat behavior, which determines the value of 
$\chi_0$, does not occur in our HLC dataset.

In terms of $\epsilon$ and $\chi_0$, there is only marginal
difference between the ``warm'' and ``cold'' HLC positions 
defined in \S3.  Separate $L^2$ analyses
for the warm and cold data points yield
$\epsilon_{\litl WARM} = 0.053{{+0.003}\atop{-0.007}}$ and 
$\chi_{0,{\litl WARM}} \gtrsim 1.0$ ($3\sigma$ lower limit);
and $\epsilon_{\litl COLD} = 0.045\pm 0.004$ and 
$\chi_{0,{\litl COLD}} \gtrsim 0.8$ ($3\sigma$ lower limit).  
Within the 1$\sigma$ uncertainties, the $\epsilon$ values are identical.  The
lower limits on $\chi_0$ are consistent with each other.  In addition, both 
the warm and cold 3$\sigma$ 
confidence contours overlap each other nearly exactly.  We conclude 
that {\it translucent cloud positions with 
warm and cold {\sl IRAS} 60/100 colors have nearly the same values of 
$\epsilon$ and $\chi_0$}.  

\begin{figure}
\plotone{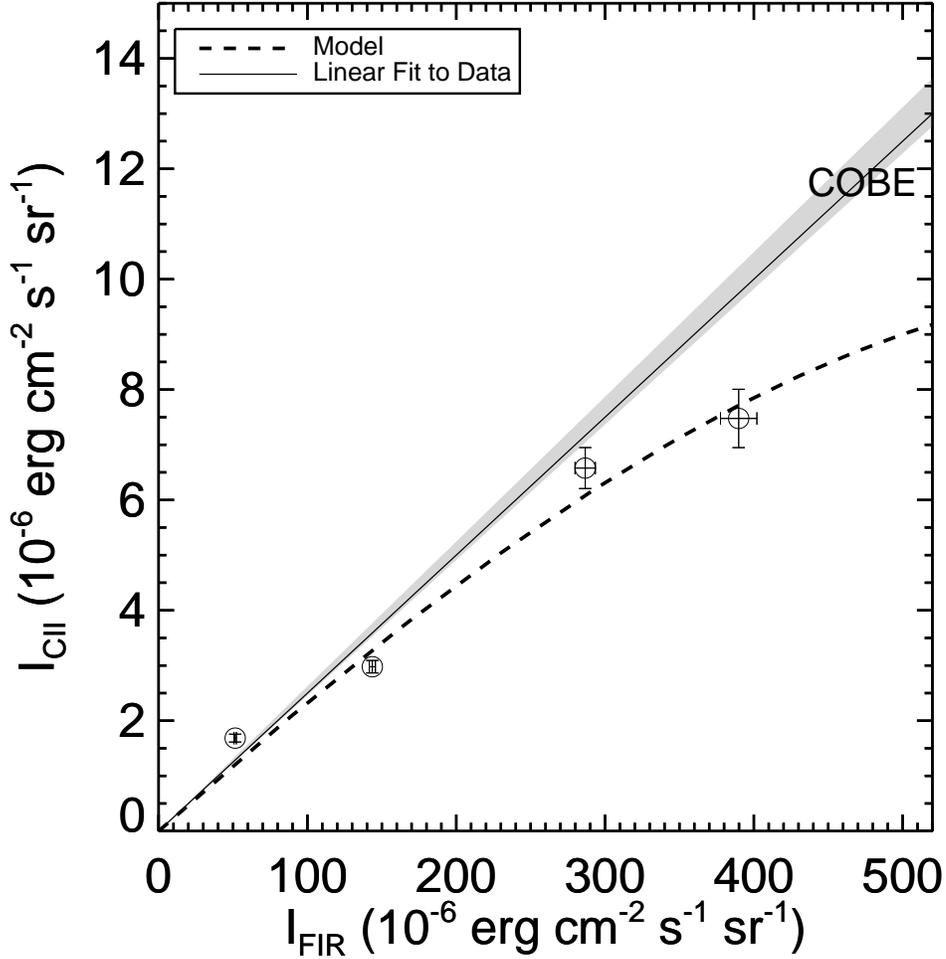}
\caption{Binned HLC measurements of \icii\ versus $FIR$ (Figure 
\ref{cp_fir_bins}), compared with model results.  The gray wedge represents
the range of {\sl COBE} data, as described in the caption to Figure 
\ref{cp_fir}.  The dashed curve represents the best fit model,
$\epsilon = 0.043$ and $\chi_0 = 1.6$.  A solid line shows the fit to 
our HLC data (Figure \ref{cp_fir}).
\label{cp_fir_bins_model}}
\end{figure}

Figure \ref{cp_fir_bins_model} displays the best fit model for 
\icii\ versus $FIR$, superimposed 
on the binned data from Figure \ref{cp_fir_bins}.  The observed decrease 
in [\cii ] emissivity at 
high values of $FIR$ (Figure \ref{cp_fir_bins}) is reproduced by the
model, signifying that HLCs with
$FIR\gtrsim 3\intint{-4}$ absorb most incident FUV photons.

\section{Discussion}

In this paper we have studied the [\cii ] and far--infrared (FIR) emission 
towards high--latitude translucent molecular clouds (HLCs).  These 
observational signatures of the processing of the interstellar radiation 
field (ISRF) by the clouds help to elucidate the role of dust grains 
in heating the gas.

\subsection{The [\cii ] emissivity and extinction}
The chief conclusion to be drawn from the new {\sl ISO} data is that,
for $FIR\lesssim 3\intint{-4}$,  \icii\ 
and $FIR$ are linearly correlated on $\sim 1\arcmin$ scales
(Figure \ref{cp_fir}), confirming the 
7$\arcdeg$ {\sl COBE} FIRAS result, $\icii \propto FIR^{0.95}$, for the 
Galactic plane \citep{ben94}.  The confirmation is not just qualitative:
 the proportionality constant is the same 
for both {\sl COBE} and {\sl ISO}.  At low extinction the 
$FIR$ integrated intensity is proportional to the \hi\ column density, so
the [\cii ] {\it emissivity} is constant for all low-extinction 
interstellar regions.

The [\cii ] emissivity decreases for high extinction lines of
sight due to the attenuation of FUV photons that cause photoelectric 
heating, an effect that was unnoticed in the low resolution
{\sl COBE} data.  When 
we bin the data (Figure \ref{cp_fir_bins}) they show a 
statistically significant decrease in slope for 
$FIR\gtrsim 3\intint{-4}$.  We 
do not have enough high extinction sources in our sample to be sure that
\icii\ flattens completely, as predicted by our model.  At the minimum, 
there is a link between the decreasing emissivity of [\cii ] and the 
transition from mostly ``warm'' to mostly ``cold'' values of \irasc/\irasd,
because most warm positions have low $FIR$ intensity and
most cold positions have high $FIR$ intensity.  

The {\sl COBE} data do not show a decrease in [\cii ] 
emissivity for high values of $FIR$, probably because translucent 
HLCs are diluted inside the 7\arcdeg\ {\sl COBE} beam.  The mean 
CO emission size of HLCs 
is $\sim$ 1\arcdeg\ \citep{mag96}, but detectable CO emission from HLCs
covers less than 3\% of the sky \citep{mag00}.  Thus on 7\arcdeg\ 
scales, $\sim 97\%$ of [\cii ] emission
is from purely atomic material, so the {\sl COBE}
survey yields only linear behavior for \icii\ versus $FIR$.  The 70\arcsec\ 
{\sl ISO} beam, on the other hand, is capable of resolving translucent
HLCs where the column-averaged [\cii ] emissivity is much lower than in
diffuse \hi .

The 60\micron/100\micron\ color is often used to infer 
either the ISRF flux \citep[e.g.,][]{dal01} or the grain size 
distribution \citep[e.g.,][]{lau91}, but our work 
indicates that for translucent clouds variations in \irasc/\irasd\
can be accounted for entirely by variations in the extinction.  Thicker 
clouds appear colder, even if 
the grain size distribution and the radiation field stay constant.  
Fitting our model to the \icii\ and $FIR$ data yields the same
values of the photoelectric heating efficiency, $\epsilon$, and the
ISRF flux, $\chi_0$, for sources with both warm and cold
60/100 colors.   

\subsection{The ISRF Flux and the Photoelectric Heating Efficiency}
The actual best fit values of the ISRF flux and the photoelectric 
heating efficiency are noteworthy.  We found that the most probable 
bolometric ISRF flux incident on HLCs is $\chi_0 = 1.6$, where 
$\chi_0\equiv 1$
defines the ISRF spectrum of \citet{mez82} and \citet{mat83}.  We also derive 
$\chi_0\approx 1.6$ {\it independently} for cloud 
MBM-12 using the observed \irasc/\irasd\ behavior as a function of visual 
extinction (Figure \ref{mbm12_av_color}).  The FUV portion of a 
$\chi_0 = 1.6$ field has an integrated flux of
$2.9\times\expo{-2}\,{\rm erg}\cmtwo\persec$.  In units of the \citet{hab68} 
flux ($1.6\times\expo{-2}\,{\rm erg}\cmtwo\persec$), this
gives $G_0 = 1.8$, close to the actual value ($G_0=1.7$) determined from 
space-based UV measurements \citep{dra78}.  Given that our model is so 
successful at determining $G_0$,
even without observing clouds in the flat regime where all FUV is 
absorbed, it is likely that the curve in Figure 
\ref{cp_fir_bins_model} is a realistic representation of the average
[\cii ] and $FIR$ properties of translucent high--latitude clouds.  This
means that the decrease in [\cii ] emissivity that we detect is indeed 
the transition 
between regions with less than complete and complete FUV absorption, 
confirming the use of the term ``translucent'' to describe the HLCs.
  We predict that the [\cii ] emissivity should decrease further for 
HLC positions with higher extinction than those in our current sample.

The photoelectric heating efficiency that we derive, 
$\epsilon = 4.3\%$, is slightly higher than the value
3\% derived from {\sl COBE} data by \citet{bak94}.  Recall that $\epsilon$
is the fraction of far-ultraviolet radiation absorbed by grains that
is converted to gas heating.  \citeauthor{bak94} 
estimated $\epsilon$ for the Galaxy by dividing the observed [\cii ] 
integrated intensity by FIR.  In so doing, they equated \icii\ with 
photoelectric
heating, as we do; but they also equated FIR with the absorbed FUV flux,
which is only partially true for translucent material.  In the cold neutral 
medium half of the ISRF 
flux that heats dust grains and causes FIR emission is at optical 
wavelengths.  The FIR integrated surface brightness 
can only be used to estimate the absorbed FUV flux if 
clouds have nearby O and B stars,
causing FUV photons to dominate the radiation field.  

Another way to obtain a crude estimate of $\epsilon$ for the 
cold neutral medium (CNM) is to consider the limiting case of complete FUV
absorption in translucent gas.  The 
[\cii ] cooling that should result is $I_{\litl CII}^{\litl max} = 1.44\times
\expo{-4}\,\chi_0\,\epsilon\,\erg\cmtwo\persec\,{\rm sr^{-1}}$ (Equation
\ref{fuv_integrated}).  For our Figure \ref{cp_fir} HLC sample, 
$I_{\litl CII}^{\litl max} \approx 1.1\intint{-5}$, so 
the product $\chi_0\,\epsilon \approx 0.076$.  Taking $\chi_0 = 1.6$ (see
above) gives $\epsilon = 4.8\%$, implying that the CNM is extremely 
efficient at processing
FUV radiation.  In fact, both the 4.8\% estimate and our best fit 
appraisal, $\epsilon = (4.3{{+1.0}\atop{-0.2}})\%$, are consistent with
the value 4.9\% adduced as the maximum 
efficiency for neutral grains \citep{wol95,bak94}.  Fewer than 50\% of the 
grains responsible for heating the CNM are expected to be 
ionized \citep{li01,wei01}, so the grains should have close to maximal
efficiency.  

\subsection{Limitations in the Grain Model}
\subsubsection{Thermal Equilibrium Assumption for Very Small Grains}

Our results depend on the assumption that all dust grains are in thermal
equilibrium.  In particular, Equations \ref{dustbalance} and \ref{dustcool}
require this condition.  As mentioned in \S4.2.1, thermal equilibrium
breaks down for grains smaller than $a\sim 250\AA$, which are
heated by individual photons.  These very small grains (VSGs) get much hotter
than the temperatures we derived by equating the average power
absorbed with that emitted.  The
effect is most obvious at short wavelengths 
($\lambda \lesssim 60\micron$).  Thus our predicted thermal equilibrium 
60\micron\ surface brightnesses are
underestimates to the actual values derived when the stochastic
heating of very small grains is considered.  According to the 
models of \citet[Figure 8]{li01}, 
stochastically heated VSGs are responsible for approximately half 
of the 60\micron\ flux from {\it diffuse} high--latitude clouds 
\citep[see also][]{ver00}.  Clearly this is an overestimate to the 
stochastic VSG contribution to the emission from translucent clouds, since 
the power absorbed (and hence
emitted) by VSGs is a strong function of the FUV flux, which decreases 
inside translucent clouds.  If our predictions for \irasc\ are 
underestimated by a factor $\lesssim 2$, then our $FIR$ predictions 
will be underestimated
by $\lesssim 20\%$ (using Equation \ref{fir_def} and the fact that 
$\irasc \approx 0.16 \irasd$ for average \hi\ clouds).  Adding 20\% to our
model $FIR$ values decreases by 20\% the slope of the 
linear \icii\ vs. $FIR$ relationship for a given ($\epsilon$,$\chi_0$)
pair.  Thus the derived product 
$\epsilon\chi_0^{-0.267}$ (Equation \ref{linear_eq}) decreases
by 20\%.  This does not change significantly our conclusions, since 
$\epsilon = 
0.043{{+0.01}\atop{-0.002}}$ is already uncertain by 23\%.   We
need to take into account the small grain physics in more detail to
assess accurately the errors due to our equilibrium approach.  This
will not change an important conclusion derived from our 
model:  variations in the [\cii ] emissivity and the 60/100 color in 
translucent clouds reflect variations in grain heating, not 
changes in grain optical properties.

\subsubsection{Variations in Grain Optical Properties}
It is likely that our conclusions will change for
extremely opaque clouds, whose interiors may 
be sufficiently dense and free of UV photons for smaller dust particles to 
coagulate into larger grains.  Two compelling pieces of evidence for 
variations in the optical properties of grains are (1) an abrupt drop in
60/100 at $\av \sim 5$ \citep[e.g.,][]{lau91,lau95}; 
and (2) far-infrared spectra indicating an extremely cold ($12-13\K$)
population of dust towards dense cores \citep[e.g.,][]{ber99,juv02}.  
Neither feature is a characteristic of our model, so we caution against
applying our conclusions directly to dark clouds.  Nonetheless, the
effects should be negligible in the model's intended domain, i.e., 
translucent clouds ($\av = 1-5$).

\subsection{Details of Cloud Structure}

In our analysis we made no explicit mention of the cloud density
structure.  We parametrized our model in terms of the visual extinction,
which allows us to ignore density structure along the line of sight.
If the volume density varies along the line of sight, the 
size scale associated with extinction varies, but this does not affect
our calculations if all else remains the same, \ie\ the photoelectric 
effect is the major heating source and [\cii ] emission is the dominant
cooling source.  This is probably true.  The brightest lines observed in
HLCs ($\icii\sim 1\intint{-5}$) account for approximately 5\%
of the incident FUV flux (when $G_0 = 1.7$; see \S5.2 and \citealt{dra78}).  
In other words, the 
FUV heating is converted to [\cii ] cooling with very high efficiency. 
This probably requires that the [\cii ] excitation 
conditions are always thermal, \ie\ either the density is above the
critical density, $\ncrit = A_{ul}/\gamma_{ul}$, where $\gamma_{ul}$ is the 
collisional
excitation coefficient; or the temperature is above $h\nu_{ul}/k = 91\K$.  
The average CNM temperature is $\sim 80\K$, but it ranges from
20--200\K\ \citep{kul87}, so the \hi\ gas is probably hot enough 
for [\cii ] to be thermalized.  For cooler gas ($T \sim 20\K$), the 
[\cii ] critical density is high.  For
excitation by \htwo, $\ncrit\approx 9000\cmthree$; for excitation by 
\hi, $\ncrit\approx 6700\cmthree$.  The 
CO-emitting molecular regions of HLCs could have such high densities 
\citep{ing00}, but it is not clear that the [\cii ]-emitting 
regions are so dense.  It is probably the case that gas which is too 
rarefied to be supercritical either is hot enough for [\cii ] to be 
thermalized or does not contribute much to the overall extinction.  
We conclude that the detailed structure along the line of 
sight therefore has little effect on our analysis.  

Structural variations
in the {\it plane of the sky} are more difficult to account for,
for two reasons.  First, clouds with structure in three dimensions are more
permeable to the interstellar radiation field, especially when one takes
into account scattering of FUV photons by dust grains \citep{spa96}.  Second, the emission
observed towards clouds is always the result of averaging the real emission
across a telescope beam.  Comparing two tracers observed
with the same beamsize towards the same position minimizes the effects of
structured emission.  Unfortunately, the {\sl IRAS} data we use 
have a resolution of 4\arcmin, whereas the {\sl ISO} [\cii ] measurements
have only 71\arcsec\ resolution---a factor of 11 in beam area!  We probably
have enough data points that structural variations averaged out in 
statistics (compare Figure \ref{cp_fir} to Figure \ref{cp_fir_bins}).
Nevertheless, to improve the robustness of our conclusions requires 
higher resolution FIR observations, perhaps made with the Space Infrared 
Telescope Facility.  

\section{Conclusions}

A linear relationship is observed between the $FIR$ and [\cii ]
emission for high Galactic latitude molecular clouds observed
with {\sl ISO} that is indistinguishable from
the prediction based on high--latitude {\sl COBE} data, implying that
the [\cii ] emissivity is constant for all low-extinction gas.  At 
high extinction the [\cii ] emissivity begins to decrease due to
the attenuation of the FUV portion of the interstellar radiation field.  
In contrast to other work, we find that differences in 
the 60\micron/100\micron\ color can be accounted for solely by extinction
variations.  Sources with 
both warm and cold colors seem to be exposed to the same radiation
field, equal to the mean field near the Sun; and have the same photoelectric 
heating efficiency, close to the value for neutral grains.  The 
transition from sources with warm to those with cold
60/100 colors coincides approximately with the transition from constant to
decreasing [\cii ] emissivity.  Such regions where the 
FUV spectrum is attenuated and softened, but is still important for 
heating the gas, define the translucent regime.

\acknowledgements
We are indebted to Laurent Cambre\'sy for providing us with an extinction
map of MBM--12, produced using his adaptive-grid star count algorithm.  This
paper benefited from the penetrating comments and helpful suggestions of 
George Helou, Francois Boulanger, and Jean-Philipe Bernard.  We thank the
anonymous referee, whose comments and suggestions helped to improve 
the final manuscript.  This work is based in part on the Ph.D. 
thesis of J. I.  We thank Dan Clemens, James Jackson, Harlan Spence, and 
Antony Stark for critical reading of an early version of the 
manuscript.  This research was supported
in part by {\sl ISO} under JPL contract \#961513.  The research described in 
this paper was carried out in part by the Jet Propulsion Laboratory, 
California Institute of Technology, and was sponsored by the National 
Aeronautics and Space Administration.

\appendix
\section{Deriving Bulk Properties of the Grain Distribution}

We assume that grains are spherical with radius $a$, and come in two kinds, 
graphite ($i=1$) and ``astronomical silicate'' ($i=2$).  The fundamental 
optical properties of interstellar grains ---absorption 
efficiency, $Q_{abs,\lambda}(a,i)$; scattering efficiency, 
$Q_{sca,\lambda}(a,i)$; and mean cosine of scattering angle, 
$g_{\lambda}(a,i)$---have been computed by \citet{dra84} and 
\citet{lao93}. The albedo can be constructed from these
basic properties, $\omega = (Q_{sca}/Q_{abs})/[1 + (Q_{sca}/Q_{abs})]$.

Following \citet*{mat77}, we assume a power law 
distribution of grain sizes:
\begin{equation}
dn_i = C_i a^{-3.5} da, ~~~~a_{\litl min} < a < a_{\litl max},
\label{graindist}
\end{equation}
where $dn_i$ is the number of grains per H atom of type $i$ with radii between
$a$ and $a+da$, and the upper and lower
size limits are $a_{\litl max} \approx 0.25\micron$ and 
$a_{\litl min} \approx 0.005\micron$, respectively.  The coefficients $C$
were determined by \citet{dra84} to be 
$C_1 = 10^{-25.16}\,{\rm cm}^{2.5}\,({\rm H~atom})^{-1}$ for graphite and 
$C_2 = 10^{-25.11}\,{\rm cm}^{2.5}\,({\rm H~atom})^{-1}$ for silicate.  The
values of the $C_i$ coefficients were derived by
fitting the theoretical extinction curve (see Equation \ref{extincurve}
below) to the observed average extinction curve for the Galaxy
\citep{sav79}. 

Bulk observables can
be computed by integrating a particular property over the distribution of 
grain sizes, $a$ and materials, $i$.  For example, to obtain the 
wavelength-dependent extinction, $A_{\lambda}$, one integrates
the extinction cross section, $\sigma_{ext,\lambda}(a,i) = \pi a^2 
Q_{ext,\lambda}(a,i)$ [where $Q_{ext,\lambda}(a,i)\equiv Q_{abs,\lambda}(a,i)+ 
Q_{sca,\lambda}(a,i)$], over the grain distribution:
\begin{equation}
\frac{A_{\lambda}}{\nnh} = 
\sum_{i=1}^{2} C_i\,\int_{a_{\litl min}}^{a_{\litl max}}\pi a^2 \, 
Q_{ext,\lambda}(a,i)\, a^{-3.5}\, da.
\label{extincurve}
\end{equation}
The extinction curve is usually expressed as $A_{\lambda}/\av$, which
requires that we multiply Equation \ref{extincurve}
by the standard relationship between column density and visual extinction 
$\nnh (\cmtwo) = 1.87\times\expo{21}\av ({\rm mag})$ \citep{boh78}.




\clearpage


\clearpage
\clearpage
\clearpage
\clearpage
\clearpage
\clearpage
\clearpage
\clearpage
\clearpage
\clearpage
\clearpage






\clearpage

\begin{deluxetable}{lrrcrc}
\tabletypesize{\scriptsize}
\tablewidth{3.75in}
\tablecaption{ISO LWS Observations of High-Latitude Clouds}
\tablehead{
Source Name\tablenotemark{a} & \multicolumn{1}{c}{$\ell$} & \multicolumn{1}{c}{$b$} & FOV\tablenotemark{b} & \multicolumn{1}{c}{PA\tablenotemark{c}} & Detected/\\
& \multicolumn{1}{c}{$(\arcdeg)$} & \multicolumn{1}{c}{$(\arcdeg)$} & $(\arcmin\times\arcmin)$ & \multicolumn{1}{c}{($\arcdeg$)} & Observed\tablenotemark{d}\\
}
\startdata
MBM--37 &   6.007 &   36.750 & $21.0\times 1.2$ & 90 & 7/7 \\
(L183)  &   5.778 &   36.987 & $21.0\times 1.2$ & 90 & 7/7 \\
         &   6.235 &   36.512 & $21.0\times 1.2$ & 90 & 7/7 \\
         &   7.172 &   37.040 & $\phantom{2}1.2\times 1.2$ & 0 & 1/1 \\

HD210121 & 56.876 & --44.461 & $\phantom{2}1.2\times 1.2$ & 0 & 1/1 \\
         & 56.107 & --44.188 & $\phantom{2}1.2\times 1.2$ & 0 & 1/1 \\
         & 56.436 & --44.147 & $\phantom{2}1.2\times 1.2$ & 0 & 1/1 \\
         & 56.511 & --44.019 & $21.9\times 1.2$ & 55 & 8/9 \\

MBM--28 & 141.755 &  39.047 & $\phantom{2}1.2\times 1.2$ & 0 & 1/1 \\
(Ursa Major) & 140.784 &  38.501 & $\phantom{2}1.2\times 1.2$ & 0 & 0/1 \\
           & 141.306 &  39.430 & $\phantom{2}1.2\times 1.2$ & 0 & 1/1 \\
           & 142.002 &  38.700 & $\phantom{2}1.2\times 1.2$ & 0 & 1/1 \\
           & 142.070 &  38.199 & $\phantom{2}1.2\times 1.2$ & 0 & 1/1 \\

MBM--12 & 159.251 & --34.483 & $\phantom{2}1.2\times 1.2$ & 0 & 1/1 \\
(L1457) & 159.151 & --33 817 & $\phantom{2}1.2\times 1.2$ & 0 & 1/1 \\
        & 159.301 & --33.501 & $\phantom{2}1.2\times 1.2$ & 0 & 1/1 \\
        & 159.301 & --34.471 & $30.0\times 6.0$ & 135 & 20/20\tablenotemark{e} \\
                & 159.401 & --34.450 & $\phantom{2}1.2\times 1.2$ & 0 & 1/1 \\
                & 159.601 & --34.501 & $\phantom{2}1.2\times 1.2$ & 0 & 2/2 \\

G225.3--66.3 & 225.274 & --66.280 & $\phantom{2}1.2\times 1.2$ & 0 & 1/1 \\
         & 225.312 & --65.413 & $\phantom{2}1.2\times 1.2$ & 0 & 0/1 \\

G295.3--36.2 & 295.311 & --36.141 & $21.0\times 1.2$ & 0 & 6/7 \\
         & 295.284 & --36.123 & $21.0\times 1.2$ & 0 & 7/7 \\

G300.1--16.6 & 300.055 & --16.620 & $\phantom{2}1.2\times 1.2$ & 0 & 1/1 \\
(Chamaeleon) & 299.674 & --16.318 & $\phantom{2}1.2\times 1.2$ & 0 & 1/1 \\
           & 299.914 & --16.799 & $\phantom{2}1.2\times 1.2$ & 0 & 1/1 \\
           & 300.250 & --16.955 & $\phantom{2}1.2\times 1.2$ & 0 & 1/1 \\
           & 300.263 & --16.767 & $\phantom{2}1.2\times 1.2$ & 0 & 1/1 \\

Stark 4 & 310.142 & --44.475 & $18.0\times 1.2$ & 90 & 6/9 \\
        & 309.578 & --45.141 & $\phantom{2}1.2\times 1.2$ & 0 & 0/1 \\
        & 309.747 & --44.579 & $18.0\times 1.2$ & 90 & 9/9 \\
        & 309.967 & --44.522 & $\phantom{2}1.2\times 1.2$ & 0 & 1/1 \\

MBM--33  & 359.150 &   36.578 & $\phantom{2}1.2\times 1.2$ & 0 & 1/1 \\
(L1780)  & 358.897 &   36.976 & $\phantom{2}1.2\times 1.2$ & 0 & 1/1 \\
         & 358.954 &   36.832 & $\phantom{2}1.2\times 1.2$ & 0 & 1/1 \\
\enddata
\tablenotetext{a}{Nomenclature for objects with $|b| > 25\arcdeg$ is that of 
the Magnani {\it et al.\/} (1996) 
catalog, with alternate names in parenthesis.  Separate positions within 
the same source are listed
separately.  Separate observations of the same position, including raster
maps, are listed together.  Nominal source ``center'' position, if one 
has been designated, is listed first.}
\tablenotetext{b}{Field of view of raster map.  Maps were typically made on a 
sparsely sampled rectangular grid.}
\tablenotetext{c}{Position angle of raster map in degrees counterclockwise from celestial North.}
\tablenotetext{d}{Number of positions detected in [\cii ] emission/Number of positions observed.}
\tablenotetext{e}{Previously published observations (Timmerman {\it et al.\/} 1998).}
\label{obstable}
\end{deluxetable}


\end{document}